# On the use of a consumer-grade 360-degree camera as a radiometer for scientific applications

RAPHAËL LAROUCHE,[1,2,*] SIMON LAMBERT-GIRARD,[1] CHRISTIAN KATLEIN,[1,3] SABINE MARTY,[4,5] EDOUARD LEYMARIE,[4] SIMON THIBAULT,[2] AND MARCEL BABIN[1]

*[1]Takuvik International Research Laboratory, Université Laval (Canada) & CNRS (France), Département de Biologie and Québec-Océan, Université Laval, Pavillon Alexandre-Vachon 1045, avenue de la Médecine, Local 2078, G1V 0A6, Canada*
*[2] Centre d'optique, photonique et laser (COPL), Université Laval, Quebec City, QC, Canada*
*[3] Alfred-Wegener-Institut Helmholtz-Zentrum für Polar und Meeresforschung, Bremerhaven, Germany*
*[4] Laboratoire d'Océanographie de Villefranche, Sorbonne Université, CNRS, LOV, Villefranche-sur-Mer, France*
*[5] Norwegian Institute for Water Research (NIVA), Oslo, Norway*
**raphael.larouche@takuvik.ulaval.ca*

**Abstract:**

Improved miniaturization capabilities for complex fisheye camera systems have recently led to the introduction of many compact 360-degree cameras on the consumer technology market. Designed primarily for recreational photography, several manufacturers have decided to allow users access to raw imagery for further editing flexibility, thereby offering data at sensor level that can be directly exploited for absolute-light quantification. In this study, we demonstrate methodologies to carefully calibrate a consumer-grade 360-degree camera for radiometry use. The methods include linearity analysis, geometric calibration, assessment of the illumination fall-off across the image plane, spectral-response determination, absolute spectral-radiance calibration, immersion factor determination and dark-frame analysis. Accuracy of the calibration was validated by a real-world experiment comparing sky radiance measurements with a co-localized Compact Optical Profiling System (C-OPS, Biospherical Instruments Inc.), which gave mean unbiased percentage differences of less than 21.1 %. Using the photon-transfer technique, we calculated that this camera consisting of two fisheyes with a 182° field of view in air (152° in water) has a limit of detection of at least 4.6 x $10^{-7}$ W·sr$^{-1}$·m$^{-2}$·nm$^{-1}$ in its three spectral channels. This technology, with properly stored calibration data, may benefit researchers from multiple scientific areas interested in radiometric geometric light-field study. While some of these radiometric calibration methods are complex or costly, this work opens up possibilities for easy-to-use, inexpensive, and accessible radiance cameras.

## 1. Introduction

Spectral radiance describes the positional, directional, temporal, and spectral nature of the light field when measured at a certain point inside a medium [1]. A wide range of scientific fields such as biophotonics, remote sensing, astronomy, oceanography, or medical imaging can benefit from light-field geometrical information. For example, the angular distribution of spectral radiance is used as ground truth to validate satellite measurements [2,3], to obtain material bi-directional reflectance properties [4–7], to assess artificial light pollution of the night sky [8–10], or visual comfort perception of architectural designs [11,12]. Collected over $4\pi$ steradians, radiance angular distributions can be used to retrieve and infer all apparent as well as several inherent optical properties (AOPs and IOPs) of different natural media (assuming that the contributions of inelastic diffusion are negligible) such as oceanic or inland waters [13–16], sea ice [17,18], atmosphere [19] or canopies [20]. Optical quantities thus obtained can be used to better understand light-matter interactions, energy budgets, biogeochemistry, and physical-biological processes.

In oceanography, angular distributions of spectral radiance are valuable to derive water optical properties properly and to understand the propagation of radiation through the medium. However, such measurements have not been taken frequently because of the associated technological challenges and the lack of commercial instrumentation. Based on the Gershun tube concept [21], one of the earliest radiance instruments consisted of photometer heads of 6.6˚ in field of view (FOV) mounted on a tilt and azimuthal control unit for measurements in discrete directions [22]. The very first camera dedicated to instantaneous capture of the complete light-field geometrical distribution was developed by Smith, Austin, and Tyler [23]. Their radiometer was made up of two assemblies – for the upwelling and downwelling light field respectively – of fisheye lenses, broadband photopic spectral filters and photographic films. The optimized versions, named RADS, had four automatically changing spectral filters and two different sensors for two distinct versions of the camera: a charge-injection device (CID) in the first version [24] and a charge-couple device (CCD) for the second [25]. Technological advances in electronics led to the development of a reduced-size version of the camera assembly, NURADS, measuring 30 cm in diameter and length [26]. This camera had a bulky form factor resulting in difficult mitigation of self-shadowing effects contributing to erroneous radiance measurements [27]. Significant improvements in miniaturization were made more recently with a radiance camera, CamLum CE600 (2011), that fitted inside a 9.6 cm diameter and 13 cm-length housing, representing a 15-fold reduction in volume compared to the NURADS camera [15]. The CE600 radiometer had a custom-built fisheye lens coupled with a high-sensitivity complementary metal-oxide semiconductor CMOS (measurements over 7 decades) and 6 spectral filters changed by a motorized wheel. At the same time, another radiance camera named RadCam was designed specifically for large intra-scene dynamic range [28]. This system with an equidistant fisheye lens and a single 555 nm spectral filter could capture 6 decades of dynamic range in one scene. A large dynamic range is required as, in natural environments such as the ocean, light levels vary up to 6 decades depending on the state of the incident light field (sun elevation, presence of clouds in the sky, etc.), the highly mobile and dynamic ocean surface, and the presence or absence of optically active constituents (phytoplankton, colored dissolved material, non-algal particles, etc.) [14,15,26,29]. Between measurements taken in the downwelling or upwelling hemisphere, radiance values can be separated by several decades due to the highly anisotropic nature of the underwater radiance angular distributions [28]. In sea ice, transmitted light varies by several orders of magnitude over few tens of centimeters.

Currently, narrow to wide-angle micro-optics are becoming increasingly popular. While miniature fisheye lenses (FOV ≥ 180˚) are challenging to design, new manufacturing tools and techniques, as well as ever-growing optical engineering skills, have led to the fabrication of small form factor fisheyes (total track length below 5 mm) [30]. It benefited the recreational photography market as some manufacturers have started to design 360-degree cameras capturing light rays coming from every direction. In 2018, Insta360® (Arashi Vision Inc.) launched the ONE omnidirectional (360-degree) device, whose technology lends it to use as a radiance camera suitable for new environmental applications due to its low cost, easy usage, wide availability, and small form factor.

An example of application is sea-ice internal measurements. In the past, researchers faced significant technical challenges when they wanted to measure optical properties in sea ice using available cumbersome radiometers [31–34]. This led to the under-sampling of various *in situ* internal AOPs, which currently limits the understanding of radiative transfer inside the medium. The 360-degree radiance camera, with its reduced footprint and self-shading, can be inserted in holes drilled inside ice to acquire the angular distribution of the light-field, from which all AOPs can be derived at a precise depth. Such a miniaturized omnidirectional assembly can also be integrated to drones or autonomous underwater vehicle (AUV) payloads, due to its light weight. In this way, directional reflectance properties (e.g. hemispherical-directional reflectance factor, HDRF) of surfaces like sea ice [6], open ocean and cloud [35], marginal sea-ice zones [36] or canopies [37,38], can be assessed in multiple spectral bands, over most upwelling hemispherical directions and on large spatial transects. Surface reflectance anisotropy is a key parameter for Earth energy-budget calculations and is used in multiple corrections/calibrations of satellite-sensor measurements [35,39].

The aim of this study is to offer an initial assessment and proof of concept regarding the camera's radiometric potential. These 360-degree cameras are not designed primarily for that purpose and could potentially respond non-linearly, present unwanted transformations of the available raw values or have insufficient sensitivity. We present the calibration methodologies used to properly transform the measures into spectral radiance as well as multiple characterizations to evaluate the radiometric quality of the measurements. The methods include calibration for in-water and in-air usage as many applications are related to the field of optical oceanography. We begin by setting out background radiometric equations before we describe all the methods involved in the calibration and characterization steps, then show key results from those methodologies, and finally offer statistical analysis of a sky radiance real-world validation experiment performed with the Compact-Optical Profiling System (C-OPS) radiometer [40].

## 2. Methods

### *2.1 Radiometric equations*

The 360-degree camera, with a diameter of 5 cm, includes two fisheye lenses of fixed 2.2 f-number that measure light over a $2\pi$-steradian sphere each. The imaging detectors are two Sony CMOS sensors with pixels covered in a repeated Bayer mosaic of four filters of three conventional RGGB. The CMOS have sizes of 1/2.3" (~7.8 mm in diagonal), 12 megapixels resolution (3456 x 3456 pixels) and an analog-to-digital converter (ADC) of 14 bits (16 384 possible values). An important feature of this camera is the availability of raw imagery. These are images minimally modified from the sensor-level capture. Raw images are available before being altered by an image signal-processor unit which performs tasks such as demosaicing, white balancing, noise reduction, color correction, etc. The images are saved in Digital Negative format (DNG, Adobe Inc.) which includes raw images as well as useful metadata. This opens up the possibility of using the Insta360® ONE camera for radiometry.

The readouts (digital number) $y_{DN,raw,i}$ in Analog-to-Digital Units (ADU) of a given detector element on the imaging detector are related to spectral radiance $L(\lambda)$ (units of W·sr$^{-1}$·m$^{-2}$·nm$^{-1}$) through the following equation [41]:

$$y_{DN,raw,i} = t_{\mathrm{int}} \cdot S_{ISO} \cdot A_i \cdot \Omega_i \cdot \frac{1}{hc} \cdot \frac{1}{q} \cdot \int_{\lambda} L(\lambda) \cdot \eta_i(\lambda) \cdot \lambda \cdot d\lambda + y_{DN,bias,i} \quad (1)$$

where the subscript $i$ refers to the spectral band, $t_{\mathrm{int}}$ the exposure time in seconds, $S_{ISO}$ the overall sensitivity or gain of the camera which is normalized based on ISO speed, $A_i$ (m$^2$) the entrance pupil area [42], $\Omega_i$ (sr) the solid angle covered by a single pixel, $h$ (J·s) the Planck's constant, $c$ (m·s$^{-1}$) the speed of light, $q$ (e$^-$·ADU$^{-1}$) the quantization step equivalence, $\eta_i(\lambda)$ (-) is the quantum efficiency and $y_{DN,bias,i}$ (ADU) the dark offset. The quantization step equivalence is the analog-to-digital conversion factor representing the digitization of electrons. Prior to every radiometric conversion, the images are subtracted by the dark offset and normalized by exposure parameters, i.e. gain and exposure time following:

$$y_{DN,i} = \frac{\left(y_{DN,raw,i} - y_{DN,bias,i}\right)}{t_{\text{int}} \cdot S_{ISO} \cdot 0.01} = A_i \cdot \Omega_i \cdot \frac{1}{hc} \cdot \frac{1}{q} \cdot \int_{\lambda} L(\lambda) \cdot \eta_i(\lambda) \cdot \lambda \cdot d\lambda \qquad (2)$$

with $y_{DN,i}$ (ADU·s$^{-1}$) being the digital numbers removed from the dark offset and 0.01 scaling $S_{ISO}$ relative to x1 instead of x100. The image is also downsampled into its three RGB components with the green band having pairs of consecutive rows averaged over the entire height. This method was selected to avoid modification by the demosaicing algorithm to the radiometric quality of each pixel. The spectral distribution of radiance is convolved with the spectral response of the camera for each waveband $i$ (combined responses of the lenses, the Bayer filters, and the detector element), hence the integral in Eq. (2). This integral is removed by introducing the effective spectral radiance $\overline{L}_i$ (W·sr$^{-1}$·m$^{-2}$·nm$^{-1}$) expressed as:

$$\overline{L}_i = \int_{\lambda} L(\lambda) \cdot S_{R,i}(\lambda) \cdot d\lambda \Big/ \int_{\lambda} S_{R,i}(\lambda) \cdot d\lambda \qquad (3)$$

where $S_{R,i}(\lambda)$ is the peak normalized spectral response $\eta_i(\lambda) \cdot \lambda$ [43]. Constants in Eq. (2) and Eq. (3) are merged into a coefficient $R_i$ (ADU/(W·sr$^{-1}$·m$^{-2}$·nm$^{-1}$·s)), corresponding to the radiance responsivity. We refer to the inverse of this coefficient as the calibration coefficient. To project the absolute-radiance calibration to every pixel forming the image sensor, a roll-off correction $g_i$ is added to the measurement equation [15]. We refer to roll-off as the decrease in radiance reaching the pixels on the imaging plane due to vignetting and modification of the optical throughput [44]. The optical throughput is a metric of how much light flux an optical system can collect and is proportional to the geometrical extent (also referred as etendue or optical invariant) $A_i \cdot \Omega_i$ (m$^2$·sr). A factor of immersion $I_i$ is also included to account for the response variation of the camera immersed in water, a consequence of the change in refractive index of the medium in contact with the external optical surface [45]. The measurement equation becomes:

$$y_{DN,i}(\theta,\phi) = g_i(\theta,\phi) \cdot \frac{1}{I_i} \cdot R_i \cdot \overline{L}_i \qquad (4)$$

with $\theta$ and $\phi$ being respectively the zenithal and azimuthal angles of radiance seen by the pixels. The immersion factor is calculated for the measurements along the lens optical axis (camera zenith). The difference between this immersion factor and those required for the other angles is then included in the roll-off measured in water for the other angles.

The RGB channels have broad band-pass filters designed to render colors as perceived by human vision, resulting in mixed spectral information within each channel. Given this characteristic and the necessity to adapt to various natural illumination conditions, we calibrated the camera for the effective spectral radiance, which represents the average radiance within each band, as shown in Eq. (3). We acknowledge that this approach introduces averaging errors (that add to the calibration uncertainties) compared to measurements obtained with narrow-band sensors. These errors are influenced by the spectral variability within the bands and the presence of out-of-band signals [46] . Despite these limitations, we consider the 360-degree camera to be a viable solution for low-cost radiometry, providing useful data for our applications.

The following sections describe the calibration and characterization methodologies of the variables involved in Eq. (4). Linearity, relative spectral response, and dark-noise evaluations are also presented. These experiments were conducted on both electro-optic systems of the camera assembly, but results for only one of them are presented. The geometric calibration as well as the roll-off assessment were performed in air and water to allow use of the camera in both media.

### 2.2 Linearity

The digital outputs of a camera can't be assumed linear with exposure parameters even in raw format. The radiometric calibration of the camera depends on its linear behavior with exposure time and ISO gain as the digital numbers are scaled by those quantities before the application of the absolute calibration coefficient as shown in Eq. (2). Moreover, deviations from linearity can result in inaccuracies in roll-off, spectral response, immersion factor, and absolute radiance calibrations. These calibrations depend on the camera's linear response, underscoring the critical importance of this characterization. Linearity can be verified either by variation of the camera-exposure parameters using a constant light source or by attenuation of the same source [47]. We used the former method. Images of a spatially uniform and stable radiance source were captured. This source consists of the output port of an integrating-sphere (Labsphere 3P-GPS-040-SF, 10.16 cm diameter, Spectraflect 98 % reflectance coating) illuminated by a fiber-coupled quartz tungsten halogen (QTH) lamp (Thorlabs SLS201L). The QTH lamp had a color temperature of 2796 K generating visible and IR light.

We tested the two types of linearity separately. First, we assessed the linearity of the sensor by varying the exposure times whiles keeping the ISO constant. Images of the light source were acquired at exposure times of 8.33, 16.67, 33.33, 66.67, 100, 200, and 500 ms, which was the only range possible without pixel saturation. Next, we verified the linearity of the analog-to-digital converter by varying the ISO gain at a constant exposure time. The ISO gain was changed between 100 and 3200, doubling the value each time.

For every exposure condition, 4 images were acquired. Linearity over the range of exposure time and ISO was verified by calculation of the Pearson correlation coefficients $r$ for each pixel individually [47], thus requiring no roll-off correction. For the Pearson coefficient, only one image (among the four acquired) was used to verify the worst-case scenario without any noise reduction. The linearity curves were computed after averaging the four images pixel-wise and then taking the average inside a 5° field of view mask.

### 2.3 Geometric calibration

Fisheye lenses generally suffer from distortion because of their wide field of view [48]. To correct the images for distortion, we performed a geometric calibration using the Matlab™ OCamCalib toolbox [49]. The algorithm uses a general Taylor series expansion function $f(\rho)$ to model the distortion as a function of the radial distance $\rho$ in pixels from the principal point:

$$f(\rho) = a_0 + a_2 \cdot \rho^2 + \cdots + a_N \cdot \rho^N \tag{5}$$

with $a_N$ the polynomial coefficients. The first-degree coefficient $a_1$ is null to satisfy the condition $\frac{df}{d\rho} = 0$ at $\rho = 0$ [49]. The radial distance is calculated using $\rho = (x^2 + y^2)^{1/2}$ where $(x, y)$ are the pixel coordinates of a hypothetical plane orthogonal to the lens optical axis with the plane-axis intersection as the origin. They are related to the distorted real coordinates $(x', y')$ by an affine transformation which considers misalignment between the optical elements and the sensor plane:

$$\begin{pmatrix} x' \\ y' \end{pmatrix} = \begin{pmatrix} c & d \\ e & 1 \end{pmatrix} \cdot \begin{pmatrix} x \\ y \end{pmatrix} + \begin{pmatrix} x_c \\ y_c \end{pmatrix} \tag{6}$$

with $\left(x_c , y_c\right)$ being the position of the principal points while $c$, $d$, $e$ are the variables of the stretch matrix for optic misalignments. A chessboard plate of known square size is used for the calibration. With the board imaged at multiple positions and orientations, the equation system that links a 3D point to its corresponding position on the image plane becomes overdetermined. The coefficients of the distortion function $f(\rho)$ are then fitted by a pseudoinverse of the matrix-like development of the equation system. Equation (5) was set to a fourth-order polynomial as suggested by Scaramuzza et al. (2006) [50]. The angle between a vector in the direction of a scene point and the optical axis is simply obtained by:

$$\tan\theta = \rho / f(\rho) \tag{7}$$

The central 6 x 6 corners of a 9 x 7 checkerboard with squares of 16 mm were kept for the calibration. The positions of the corners were detected using the Python OpenCV [51] *findChessboardCorners* routine as it was found to be more robust in different lighting conditions than the Matlab™ built-in algorithm *detectCheckerboardPoints*. As mentioned earlier, the calibration was performed in air as well as in water using a water tank and for the three channels separately. Inside water, a total of 20 images of the target in different positions were taken corresponding to 720 calibration points. Sixteen images were acquired for the air calibration as these provided sufficient precision. Attention was paid to cover the entire FOV to ensure validity of the inferred intrinsics (polynomial coefficients, stretch matrix variables and principal point positions) at the limit of the FOV.

### *2.4 Roll-off*

Multiple methods are proposed in the literature to characterize radiance roll-off. The most common ones make use of plain white paper [52], integrating spheres covering a large part of the optic FOV [43,47], or spectralon targets with camera rotation [15,24]. We followed the later methodology with a slight modification of replacing the spectralon by an integrating-sphere. The method solely requires rotation of the camera in such a way that the constant and spatially uniform source is imaged at different positions across the complete circle image of the sensor plane. By monitoring the average digital value of the imaged source, we observe its decline as a function of the field angle. We used the same spatially uniform and stable radiance source described in Section 2.2 for the experiment. For the in-water experiment, the 2.54 cm-diameter integrating-sphere output opening was placed in contact with one of the external walls of the aquarium (see Fig. 1a). Images were captured every 2˚ for camera rotations between -70˚ and 70˚ in water, and from -80˚ to 80˚ with increments of 5˚ for the in-air characterization. The acquired digital numbers were dark subtracted, then averaged inside a circular mask of 15 pixels in radius centered on the source spot (for noise reduction). We normalized the values by the curve's maximum. The azimuthal isotropy of the roll-off was verified by repeating the manipulations for 0˚ and 90˚ azimuthal planes.

### *2.5 Relative spectral response*

Spectral response can be characterized using light beams generated from single [43,53,54] or double monochromators [47,55,56]. We used the double-grating excitation spectrometer of Horiba Fluorolog®-3 in FL3-22 configuration. Figure 1(b) shows the schematic of the characterization experiment. Two 2.5 mm-thick Edmund Optics opal diffusing glasses (100 x 50 mm) positioned in front of the camera spread the collimated beam to avoid concentrating all the light on a small number of pixels (see Fig. 1(b)). Depending on the incident beam spatial uniformity, these diffusers can create near-Lambertian sources when used in transmission [57] and have wavelength dependencies below 5 % across the visible spectrum. To correct for the spectrofluorometer Xenon lamp spectral variations, a reference photodiode – located just before the sample compartment – monitored the beam's intensity. Based on an experience with a similar instrument (Fluorolog-2) and the use of a chemical quantum counter (oxazine; [58]), spectral relative difference between light received by the reference photodiode and that reaching the sample compartment (effects of beam splitter and two surface mirrors) is smaller than 2% in the visible.

We started the experiment by acquiring approximately 250 intensity points of the 1 nm-slit beam with the reference photodiode at an exposure time of 0.1 s. Simultaneously, we captured four images with the radiance camera at a gain of 100 and an exposure time of 0.1 s. The images were then dark corrected and pixel-wise averaged. Further noise averaging was accomplished using pixels inside a 1˚ mask around the centroid of the beam. We repeated this procedure for wavelength increments of 10 nm between 400-700 nm. Each spectral digital number was normalized by the beam intensity at the specific wavelength. The curves were finally divided by their peak value to obtain the relative spectral radiance. We computed the effective wavelength from the following relationship:

$$\lambda_{eff_i} = \frac{\int \lambda \cdot S_{R,i}(\lambda) \cdot d\lambda}{\int S_{R,i}(\lambda) \cdot d\lambda} \tag{8}$$

while the bandwidth of each spectral band was found from $\Lambda_i = \int S_{R,i}(\lambda) \cdot d\lambda$. We used the trapezoidal rule for numerical integrations.

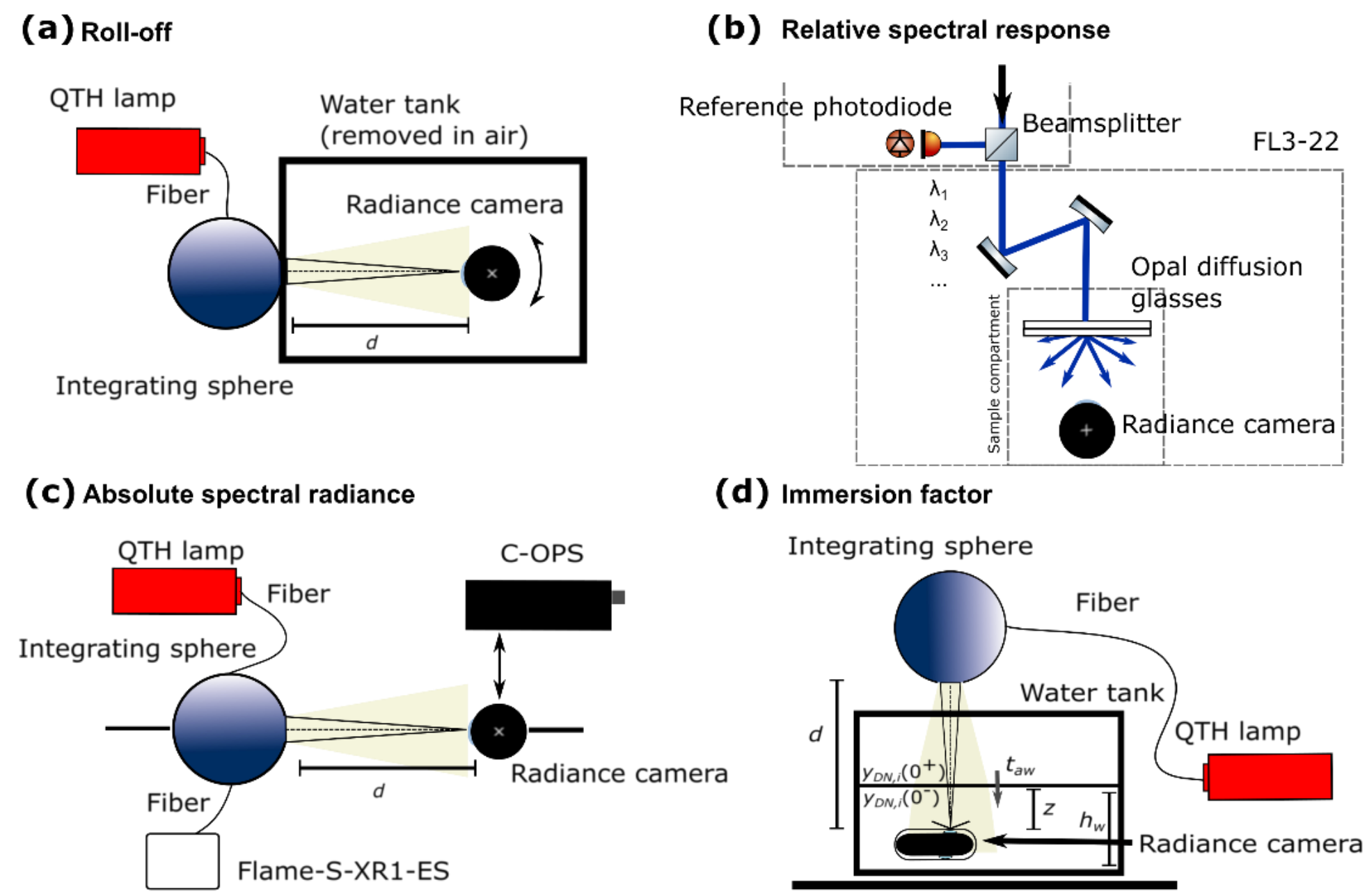

Fig. 1. Experimental setups for (a) roll-off calibration (view from above) in air and in water – the latter executed inside a tank filled with water –, (b) relative spectral response obtained from usage of Horiba Fluorolog®-3 spectrofluorometer in FL3-22 configuration, (c) absolute spectral-radiance calibration (view from above) and (d) immersion factor (side view). The calibration setups for the roll-off, the absolute-spectral radiance and the immersion factor all included a quartz tungsten halogen (QTH) lamp fiber-coupled to an integrating sphere.

## *2.6 Absolute spectral radiance*

A source of known spectral radiance $L_{source}(\lambda)$ is needed for the absolute calibration. From this information, the effective radiances $\bar{L}_{i,source}$ are calculated using Eq. (3). The ratio of these values to the digital outputs generated by the source – dark subtracted and normalized by exposure parameters – give the inverse radiance responsivity $R_i^{-1}$:

$$R_i^{-1} = \frac{\overline{L}_{i,source}}{y_{DN,i,source}} \tag{9}$$

Conventional radiometric absolute calibrations make use of an expensive NIST-calibrated FEL lamp with a reflecting spectralon plaque [15,24,26]. Instead, we used the isotropic source described in Section 2.2. We characterized the spectral radiance emitted from the 1-inch (1 in. = 2.54 cm) port using two measurement units: a fiber optic spectrometer (FLAME-S-XR1-ES, Ocean Optics Inc.) for spectrum determination and the C-OPS radiometer (calibrated 19 months earlier) [40] for absolute spectral radiance scaling. This scaling was achieved by applying a factor to the measured spectral distribution that matched the value at 589 nm to the radiance of the C-OPS measured at this same wavelength. The spectrometer was calibrated using Ocean Optics Inc. HL-3P-INT-CAL lamp (calibrated in irradiance 7 months prior) as recommended by the manufacturer. A schematic of the calibration experiment is shown in Fig. 1(c). After characterizing the light source spectral radiance, we replaced the C-OPS by the camera and 10 images of the light source, and 5 dark frames were taken. Each image stack was averaged pixel-wise for noise reduction. The digital values inside a mask of $\leq 5^\circ$ around the optical center were averaged for further noise reduction while keeping the angular range low enough so that illumination fall-off would be negligible. The experiment was conducted in a dark laboratory.

### *2.7 Immersion factor*

The theoretical equation for the immersion factor adopted by the ocean optic community is expressed as follows:

$$I = \frac{n_w \cdot (n_w + n_g)^2}{(1+n_g)^2} \tag{10}$$

with $n_w$ the refractive index of water and $n_g$ the refractive index of the external surface [45]. We used the procedure described in Zibordi (2006) to experimentally characterize the immersion factor [45]. Compared to their original setup, we inverted the detector and source positions. A schematic of the experiment is shown in Fig. 1(d). The immersion factor $I_i$ is found following:

$$I_i = \frac{y_{DN,i}(0^+)}{y_{DN,i}(0^-)} \cdot t_{aw} \cdot n_w \tag{11}$$

with $y_{DN,i}(0^+)$ and $y_{DN,i}(0^-)$ being the digital measurements when the camera is above and in water respectively, $t_{aw}$ the transmittance at the air-water interface and $n_w$ the refractive index of water [45]. These two last variables correct for the increase in radiance below water. The experiment was conducted using Milli-Q water to ensure repeatability and usage of known values for $t_{aw}$ and for $n_w$. At first, the in-air response was captured in an empty tank. The $y_{DN,i}(0^-)$ value was estimated with the intercept of a linear regression of the log-transformed digital measurements as a function of water level $z$ over the camera. This fitting method reduces the uncertainties due to source anisotropy [45]. Water levels $h_w$ (Fig. 1(d)) were varied between 6 and 11 cm by 0.5 cm increments. Five images were acquired at each water level for noise reduction. We calculated $y_{DN,i}(0^+)$ and $y_{DN,i}(z)$ for the three spectral channels using the average of pixel values inside a mask of $1.5^\circ$ around the central point. This small angular mask prevents the effect of roll-off attenuation and ensures remaining within the small-angle regime of the transmittance.

### 2.8 Dark frame

Dark frames are a combination of an offset bias and sensor-level noises such as reading and thermal noises. Offset bias may vary from one pixel to another resulting in a spatially fixed pattern noise (FPN). For the noise which adds to the bias, the level depends on the selected gain, the exposure time and the sensor temperature. As the sensor temperature was unavailable, we did not test the temperature dependency. As for the linearity method in Section 2.2, the ISO gain was varied between 100 and 3200 by doubling the value at each step. Exposure time was changed over 5 orders of magnitudes between 0.25 ms and 2000 ms. The frames were taken in a dark laboratory at a temperature of approximately 20˚C with a black absorbing fabric covering the camera. A total of 15 frames for each combination of ISO and exposure time was acquired.

### 2.9 Real-world validation

To verify the precision and accuracy of the radiometric calibrations, we carried out a comparison between the sky spectral radiances measured with the Insta360® ONE camera and the Compact-Optical Profiling System (C-OPS) radiance sensor (calibrated in February 2019). We chose to take the measures at the end of the day as the sunset allowed more dynamic range. On 18 March 2021, the C-OPS was placed on a tripod in the middle of an open field on the campus of Université Laval in Quebec City (46˚46'48.1" N, 71˚16'31.3" W). It was oriented to point upward toward the zenith. The radiance camera was fixed approximately 10 cm away from the radiometer with the optical axis of one of the fisheye lenses also oriented toward the zenithal direction. The sky conditions were clear that day with few clouds and the temperature was around 2˚C. We started the experiment at 16h00 (Eastern Daylight Time, North America) and it lasted 1 hour as the camera failed because of battery shortage. The sun elevation during that hour decreased from 27.5˚ to 18.5˚. The C-OPS was set at an acquisition rate of 6 Hz while images were acquired manually every 2 minutes with the radiance camera.

The pixel digital values were transformed into calibrated spectral radiance using the inverse of Eq. (4). We then calculated the average for each band of the camera using a 9˚ mask around the principal point, angle which corresponds to the in-air half FOV of the C-OPS microradiometer [40]. With the timestamp of both the C-OPS data and the camera images (in DNG metadata), we temporally matched the radiance values from both instruments and calculated the mean unbiased percentage difference (MUPD):

$$MUPD_i = 2 \cdot 100 \cdot \frac{1}{N} \sum_{n=1}^{N} \left| \frac{L_{COPS,i}(t_n) - \bar{L}_i(t_n)}{L_{COPS,i}(t_n) + \bar{L}_i(t_n)} \right|, \tag{12}$$

with $L_{COPS,i}$ and $\bar{L}_i$ representing respectively the C-OPS and camera spectral-radiance measurements for each band $i$ and every co-localized timestamp $t_n$ between $n=1$ and $n=N$. In order to obtain proper comparisons, we chose the C-OPS spectral channels closest to the radiance camera spectral bands' effective wavelength characterized in the experiment described in Section 2.5.

## 3. Results and discussion

### 3.1 Linearity

Pixel-wise linearity for a single acquisition at each exposition time was evaluated for 24 866 pixels inside a 5˚ mask around the principal points. The Pearson coefficients of linearity for ISO (gain) and exposure time were computed for each pixel and are presented as histograms in Fig. 2. The CMOS sensor had highly linear behavior with average coefficients $r$ of 0.9992 and 0.9961 for exposure time and gain, respectively. Figure 3 shows the mean digital outputs as a function of exposure time and ISO gain for the three bands separately. Linearities for the exposure times and gains presented in Fig. 3 are higher ($r \geq 0.99996$) than the ones in Fig. 2 because of the multiple frames averaged for noise reduction. The linearity holds up to saturation $y_{DN,sat,i} = 15576 = 2^{14} - y_{DN,bias,i}$ reached by the red and green bands ((a) and (b) of Fig. 3, respectively) because of their larger sensitivities and the absolute spectral radiance of the source. Linearity is higher for variations over exposure time compared to gain as the latter also amplifies noise. Nonlinearities were assessed using the residuals of the fitted curves (see Fig. S3 in the supplementary document). For exposure time, they are less than 4 % for pixels above 100 ADU. Nonlinearity in the blue band reaches 18.2 ± 38.1 % at 55.4 ADU, observed with an exposure time of 8.33 ms. This deviation from linearity may be overestimated due to high noise levels and the uncertainty associated with the fitted line, as reflected in the total uncertainty of 38.1 %. As apparent for gain linearity (bottom row of Fig. 3), the digital number at an ISO/100 of 1.0 seems to also be affected by nonlinearity effects compared to the other digital outputs. These nonlinearities range from 23.9 ± 14.2 % to 25.1 ± 13.8 %, whereas for other gains, they remain below 1.5 % (see Fig. S3). Nonlinear responses are more prevalent in CMOS imaging detectors compared to their CCD counterparts [59]. If possible, users should always set a fixed ISO gain and then vary the exposure time to achieve better a signal-to-noise ratio (SNR).

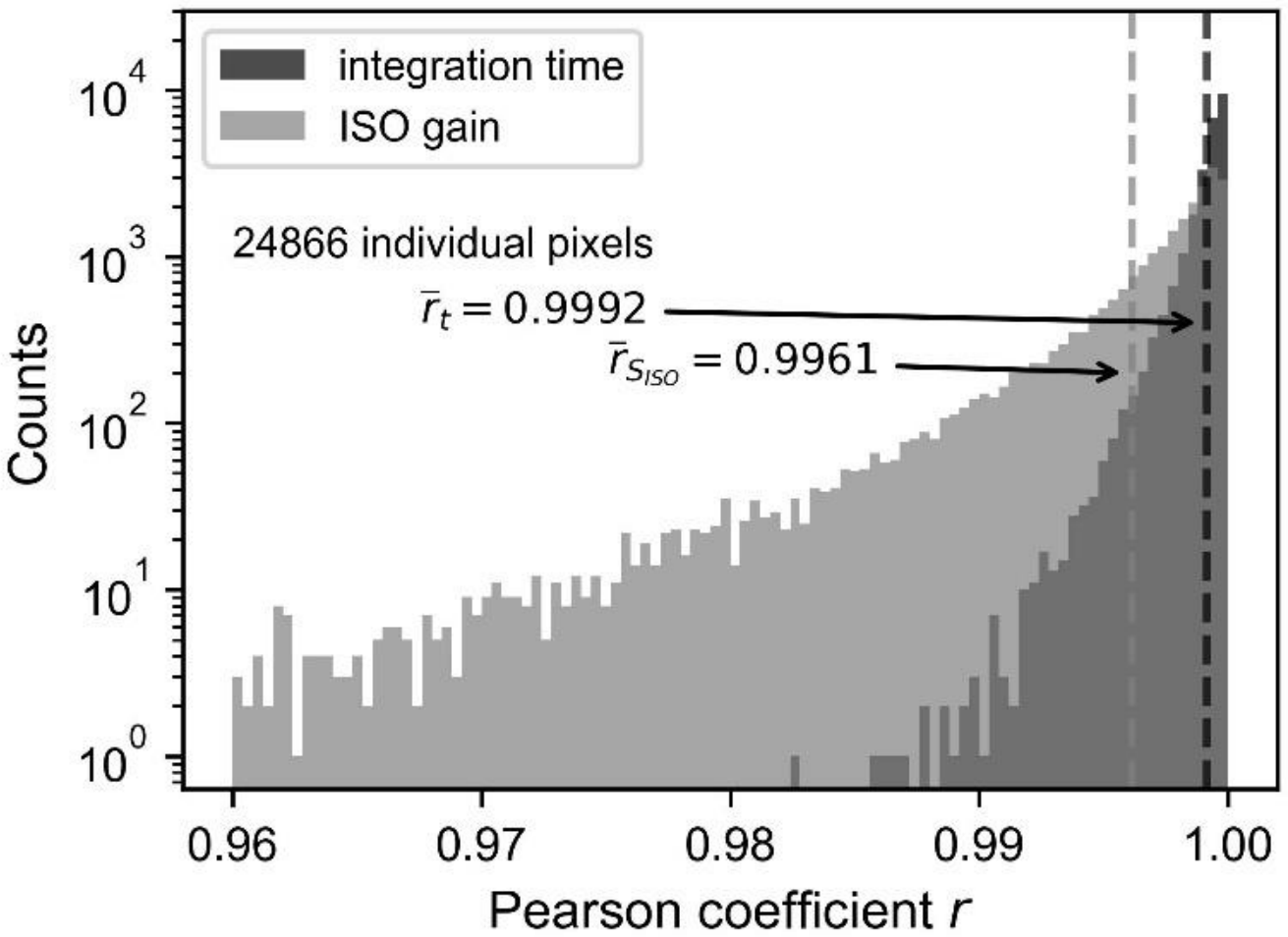

Fig. 2. Histograms of Pearson coefficients $r$ for linearity of exposition time and gain of 24 866 pixels. In both cases, pixels show linear behavior. The average $r$ for linearity in exposure time $t$ and ISO gain $S_{ISO}$ are respectively 0.9992 and 0.9961.

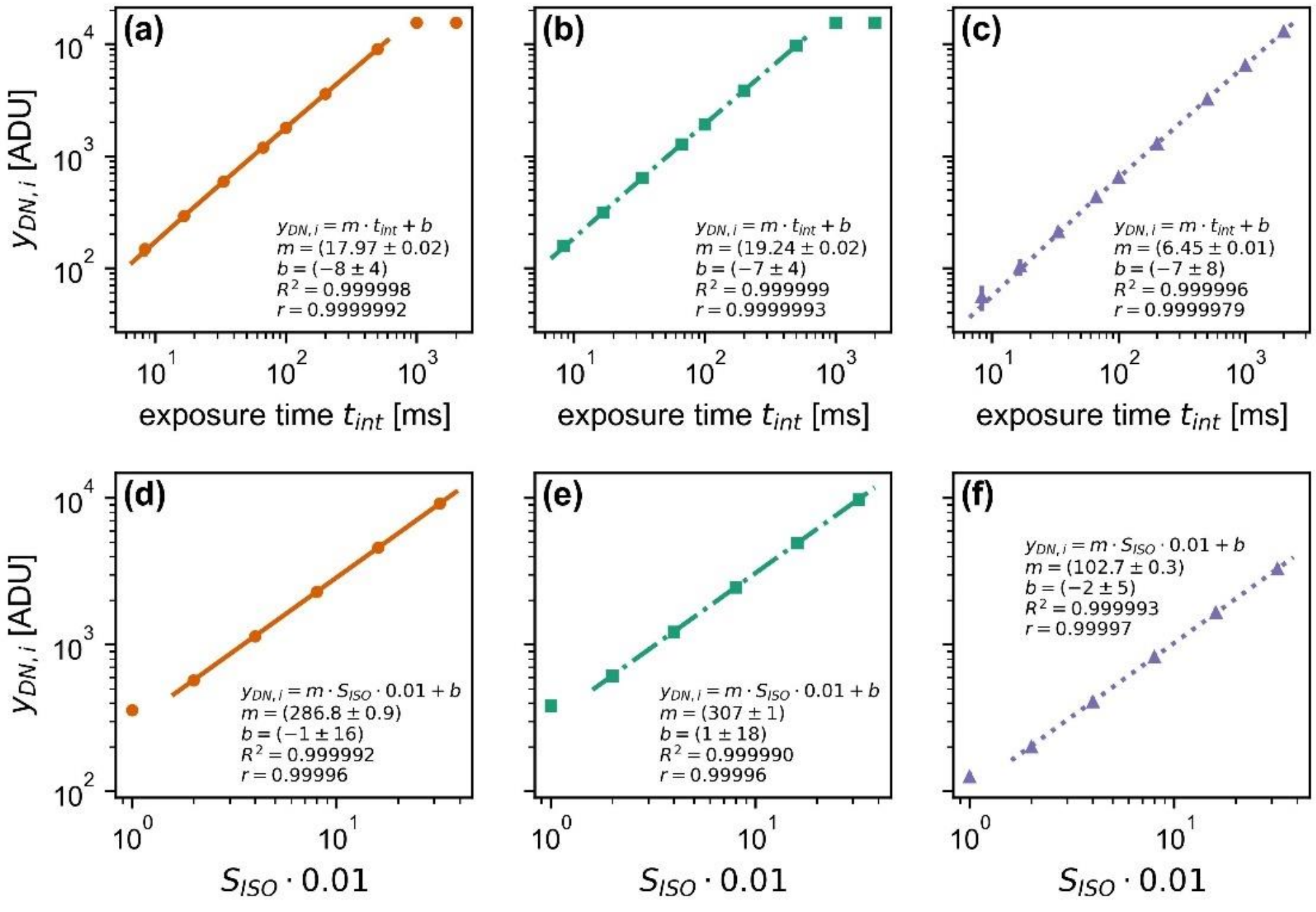

Fig. 3. Linearity of the camera over exposure time in milliseconds (upper row) and ISO gain $S_{ISO}$ x 0.01 (bottom row). From left to right, the figures show the results for the red (a)-(d), green (b)-(e), and blue (c)-(f) bands, respectively. The linear-regression estimators are given with their uncertainties corresponding to confidence intervals for 95 %. The error bars (not visible for each data point) are the standard deviations of the pixel inside the averaging mask.

### *3.2 Geometric calibration*

Viewing angles as a function of pixel Euclidean distances $\rho$ are given in Fig. 4(a) and Fig. 4(b) for in-air and in-water geometric calibration respectively. The square markers in the graphs show the direct relationship between the angles of the points in the scene (relative to the optical axis) and their corresponding positions on the sensor, i.e. before any geometric correction is applied. In those figures, we also display common stereographic and orthographic fisheye projection curves as references. It appears that the radiance camera mapping function is close to equidistant f-theta projection, meaning that the resolution in pixels per degree is nearly constant over the entire field of view. Figures 4(a)-(b) present solely the calibration results for the green spectral channel, while Fig. 4(c)-(d) (air and water respectively) show the absolute differences in the mapping function of the other bands compared to the green channel. The maximum departures are at the edge of the FOV and are of -0.27˚ (in air) and 0.52˚ (in water). This suggests that the chromatic aberrations are well corrected. As the camera was designed mainly for in-air utilization, higher chromatic aberrations were expected in water. The reprojection error is an indicator of the geometric calibration accuracy. In air, the average deviation in pixels (degrees) between the positions of the corner-algorithm extracted and the re-projected points are 0.36 px (0.026˚), 0.29 px (0.023˚) and 0.34 px (0.026˚) for the red, green, and blue wavebands. In water, they are respectively 1.59 px (0.094˚), 1.34 px (0.076˚) and 1.78 px (0.102˚). The errors are slightly higher in the red and blue channels. This may have resulted from the lower level of signal, thus lower SNR for those bands because of their narrower spectral bandwidth, which decreased the performance of the corner-detection algorithm. Figure S1 in the supplemental document shows the reprojection errors as a function of the radial distance from the principal points.

The FOV was estimated by fitting the best radial distance from the principal points to the edge of the circular images. The canny edge detector algorithm was used to generate the borders of the fisheye circular image. From the edge image, we fitted the best circle using a Circular Hough Transform. The isolated radius was used with the geometric calibration to retrieve the angle at the limit of the circular image, giving the FOV. Both algorithms were part of the Python image-processing toolbox scikit-image [60]. Using 10 different images, each waveband gave the same half field of view (HFOV) of 76˚ (total field of view of 152˚) in water. This reduction in field of view conforms to expectations as there is an optical power loss for the first lens surface due to the lower refractive index differential between water-plastic compared to air-plastic. In the air, the fitted HFOV is 91˚ for a full field of view of 182˚.

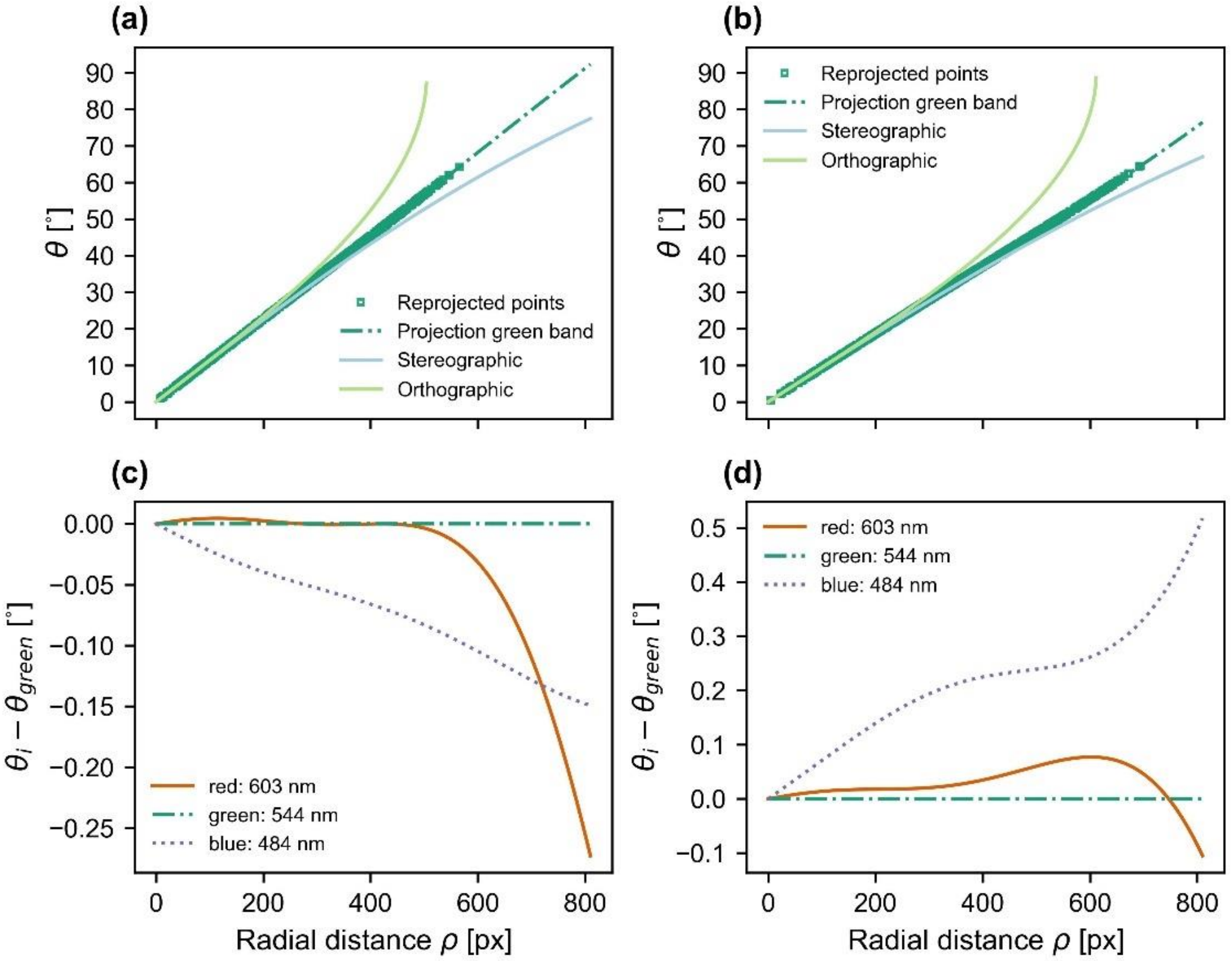

Fig. 4. Geometric calibration results. The left and right columns present respectively the calibration achieved in air and in water. The two graphs in the first row (a)-(b) show the projection function between a scene point and its corresponding radial position on the sensor. Other commonly found projection functions for fisheye lenses were also added for comparison (stereographic and orthographic curves). Only the green band calibration is shown as the spectral differences were not distinguishable. The absolute differences of the red and blue channels with the green curves are shown in (c) and (d). The spectral discrepancies are all below 0.52˚ across the entire image plane.

### *3.3 Roll-off*

Relative fall-off of illumination for off-axis points on the image plane is shown in Fig. 5. For each medium and spectral band, the measurement points shown in Fig. 5 include the two azimuthal planes $\phi$=0˚ and $\phi$=90˚. The azimuthal plane measurements were merged as no significant discrepancies were found between them. The acquisitions between -80˚ and 0˚ (or -70˚ and 0˚ for water) were converted into absolute angles using the angular calibration results of Section 3.2. The relative illumination (RI) for each spectral band shows good symmetry in the absolute angles (see Fig. 5). An 8$^{th}$-degree polynomial function of even terms was used to fit the observations:

$$g(\theta) = a_0 + a_2 \cdot \theta^2 + a_4 \cdot \theta^4 + a_6 \cdot \theta^6 + a_8 \cdot \theta^8 \quad (13)$$

with $a_0, a_2, a_4, a_6, a_8$ being the polynomial coefficients. The resulting curves are given in Fig. 5 with their respective coefficient of determination, while the fitted 8[th]-degree polynomials are provided in the supplemental document (see Table S2). The in-air fall-off at 80˚ are 0.645, 0.637 and 0.577 for the blue, green, and red bands respectively. For the field angle of 70˚ in water, the results are 0.570 (0.728 in air), 0.562 (0.719 in air) and 0.491 (0.669 in air) for the blue, green, and red pixels. For each roll-off measurement, we used the standard deviation of the pixels contained in the region of averaging as uncertainty. The mean relative uncertainties over all the angles for the red, green and blue channels are respectively 3, 2 and 4 % in air. These same values increase to 8, 5 and 12 % in water.

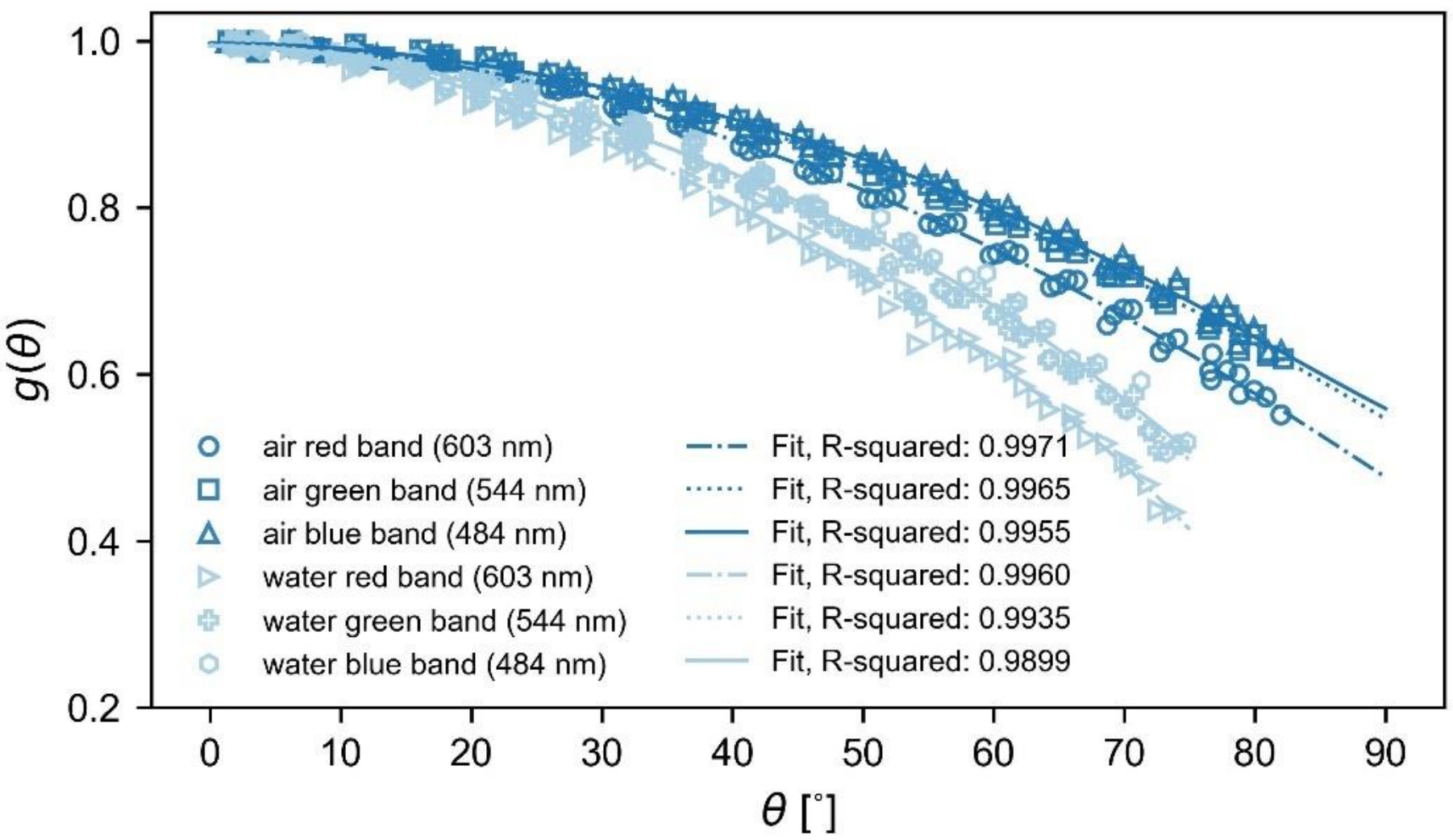

Fig. 5. Roll-off measurements as a function of angle $\theta$ relative to the camera optical axis for characterizations inside air and water. The graph shows the measurements for each band and both media as well as their respective fit using 8[th]-degree even-term polynomials. The full, dotted and dashed-dotted lines fit the blue, green, and red spectral channels in order. The coefficients of determination are given for each fit.

Illumination fall-off curves inside water have almost the same shape as their in-air counterparts but are compressed toward smaller angles. The RI at a certain pixel position will be determined by the entrance pupil area $A_i$ (affected by pupil aberrations, vignetting) and solid angle $\Omega_i$ (affected by distortion) as seen by that same element [42]. As we move away from the optical center, the geometrical extent $A_i \cdot \Omega_i$ reduces. For one pixel position on the image plane, an optical ray comes from a larger object-space angle for a system immersed inside air compared to water in accordance with Snell's law. This largely explains why the same level of $g(\theta)$ occurs at a greater angle in air than in water (see Fig. 5). The dependence of the illumination roll-off on the medium is largely determined by the projection function (distortion), which may be also affected by different levels of in-air versus in-water aberrations. The dependent aberrations of those mediums can further modify pupil area, which ultimately has an impact on the RI. Also, for rays ending at the same position on the sensor but from different incoming angles that depend on the external medium, Fresnel reflection-loss discrepancies at the interface may be significant enough to affect the roll-off [61]. Values of $g(\theta)$ are the same for the blue and green spectral bands, but decrease more strongly for the red. As shown in Fig. 4(c)-(d), the spectral differences between the projective functions are below 0.52˚. Thus, the spectral roll-off discrepancies could not have resulted from chromatic lateral color only. However, larger pupil aberrations and vignetting in the red spectral band reducing $A_i$ may explained those divergences [42].

### 3.4 Relative spectral response

Spectral responses normalized by their peak value are presented in Fig. 6 for the red, green, and blue wavebands. The results of $\lambda_{eff} \pm \Lambda_i/2$ for the red, green, and blue wavebands are 600 ± 38 nm, 540 ± 60 nm, and 480 ± 49 nm. As expected, the spectral responses are broad and overlap each other. These result mainly from the combination of the spectral responses of the filters in front of the pixels and those of the pixels themselves. The green band has the largest bandwidth of ~120 nm and will generate a larger number of counts given a spectrally constant incident-light field. The shaded region around the curves in Fig. 6 was obtained by a propagation of uncertainties related to the spectrofluorometer beam spectral energy and the digital numbers measured inside the averaging region. For signals over 10 % of the peaks, the maximum uncertainties are respectively 5.5, 4.5 and 4.5 % for the red, green, and blue pixels. Larger uncertainty for the red band may be explained by the smaller bandwidth of the channel coupled to an important decrease of the beam flux in that spectral region, both of which contribute to a reduced signal-to-noise ratio.

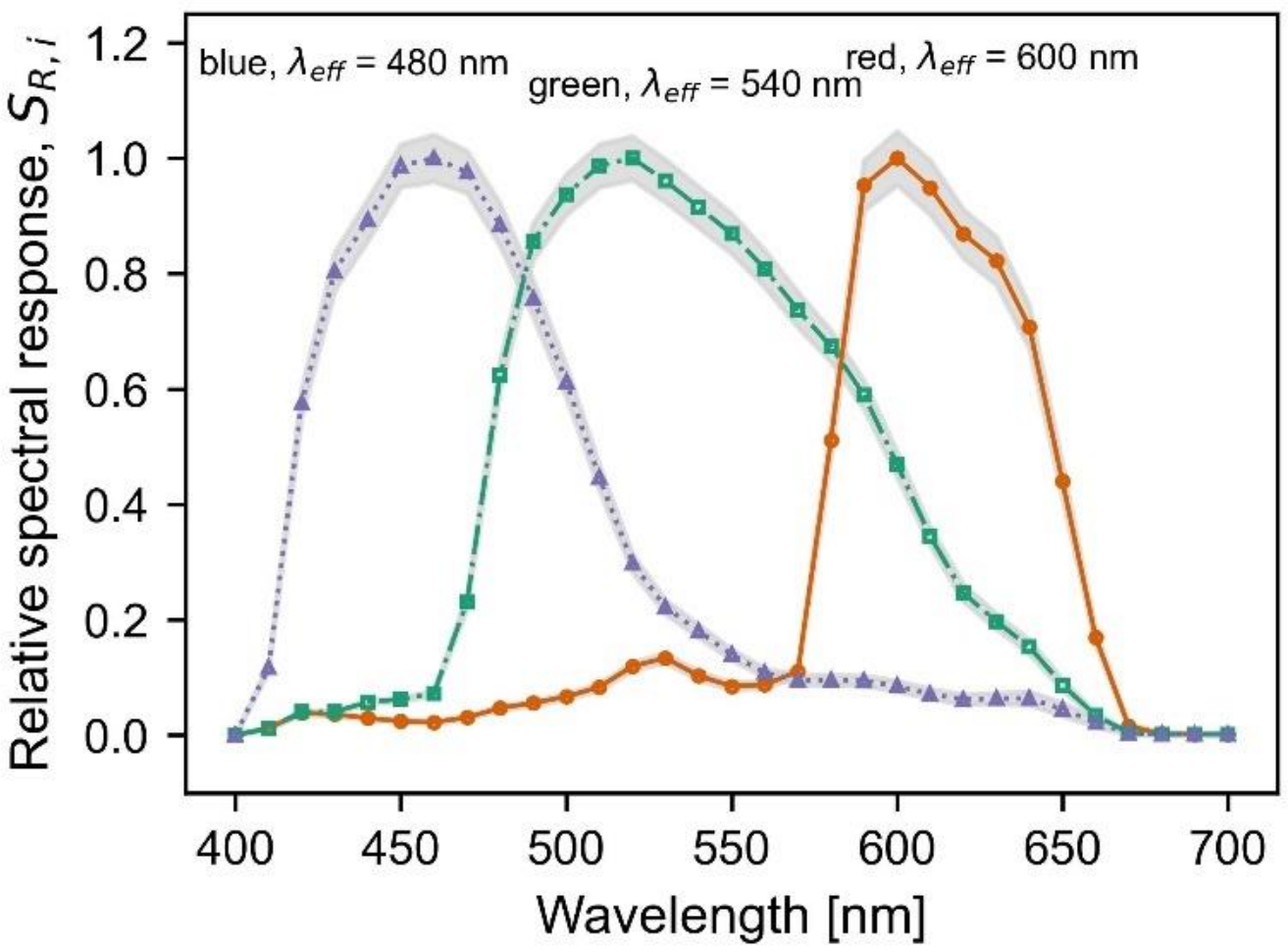

Fig. 6. Relative spectral response for the three channels of the radiance camera measured with a Fluorolog®-3 spectrofluorometer. The spectral range of the measurements is between 400-700 nm. The shaded regions present the uncertainties for every spectral measurement.

### 3.5 Absolute spectral radiance

Table 1 gives the absolute spectral-radiance calibration coefficients and the variables involved in their calculations. The images of the integrating-sphere output are shown in the supplemental document (see Fig. S2). The coefficients $R_i^{-1}$ should not, in theory, have dependencies on the calibration-source spectral radiance. $\overline{L}_{i,source}$ and $y_{DN,i,source}$ are both proportional to $\int L_{source}(\lambda) \cdot S_{R,i}(\lambda) \cdot d\lambda$, which make the latter cancel out using Eq. (9). With the calibration coefficients proportional to the reciprocal of $\int S_{R,i}(\lambda) \cdot d\lambda$, the multiplication between $y_{DN,i}$ and $R_i^{-1}$ gives the convolution of $L_{i,scene}(\lambda)$ with the spectral bands of the camera as computed from Eq. (3). For broad spectral responses, significant differences between measured $\overline{L}_{i,scene}$ and the true spectral radiance at $\lambda_{eff}$ may exist as $\overline{L}_{i,scene}$ represents the spectral average of radiance within each band. Such errors are greater for unsmooth radiance spectra inside the spectral channels [46], which is certainly happening in natural scenes.

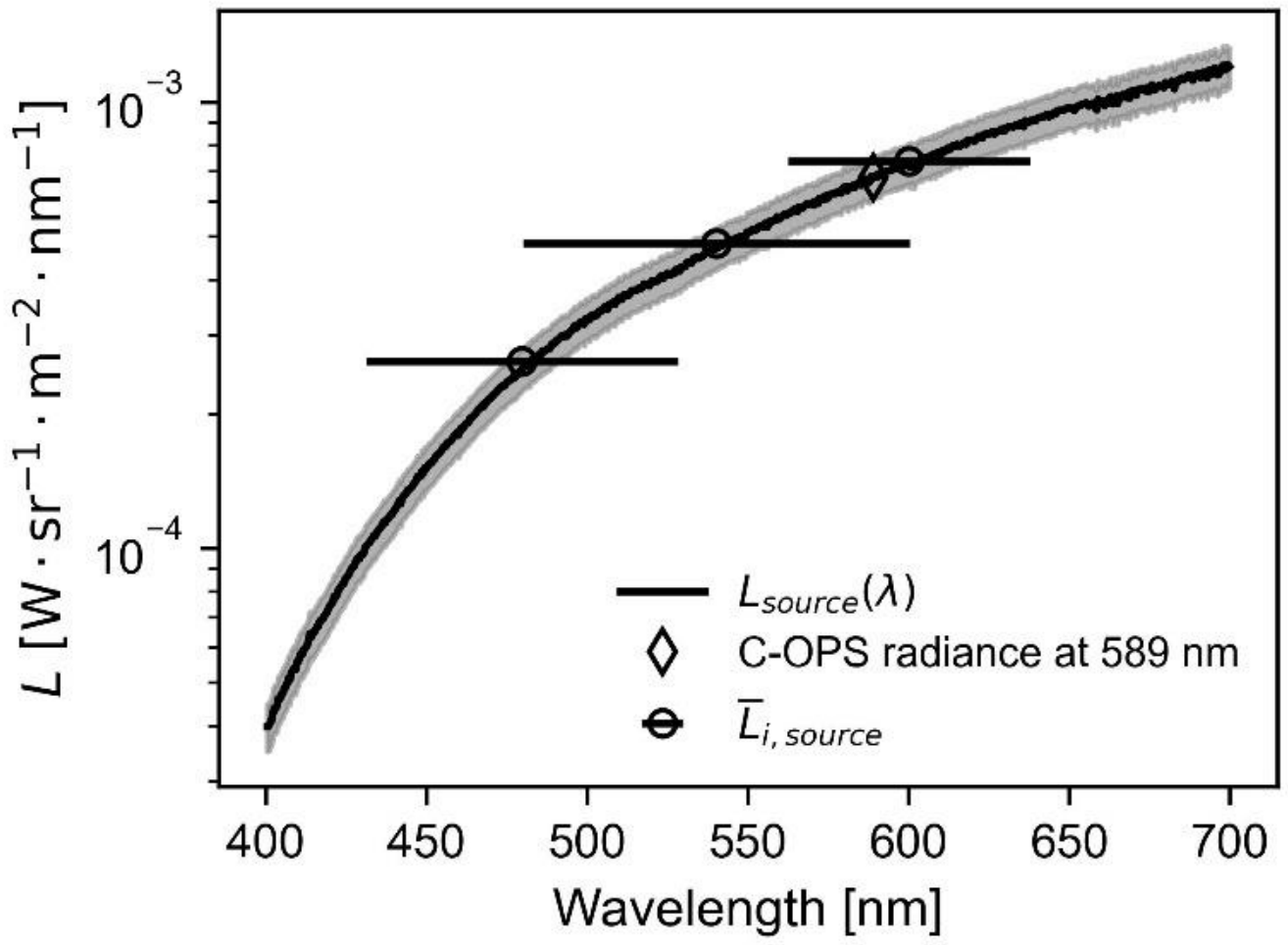

Fig. 7. Calibration-source (QTH lamp) spectral-radiance distribution. The circular markers show the spectral convolution of the source radiance with the camera channel using Eq. (3). These effective spectral-radiance values are plotted at the effective wavelength $\lambda_{eff}$ of each channel with the horizontal bars giving their bandwidth. The triangle marker represents the 589 nm-band C-OPS radiance value used to scale the source spectral flux.

The uncertainties relating to $R_i^{-1}$ are 10.4, 10.6 and 11.0% for the red, green, and blue channels, respectively. They were calculated by error propagation using Eq. (9). Uncertainties relating to digital numbers $y_{DN,i}$ correspond to the standard deviations of the averaging region. The latter uncertainties were at the level of photon shot noise. Spectral-radiance $\overline{L}_{i,source}$ uncertainties (shaded region in Fig. 7) are a combination of Ocean Optics HL-3P-INT-CAL absolute calibration lamp errors, C-OPS uncertainties, and Ocean Optics FLAME-S-XR1-ES spectrometer noise levels. The largest contributor is the Ocean Optics calibration lamp as the uncertainties given by the manufacturer are 9.1 and 6.9 % at 400 and 800 nm, respectively. Spectral radiance at 589 nm measured with the C-OPS is $0.68 \pm 0.02$ mW·sr$^{-1}$·m$^{-2}$·nm$^{-1}$ as shown by the diamond in Fig. 7. A conservative uncertainty of ±2.5 % was given to this measurement. C-OPS radiometers are commercial off-the-shelf (COTS) instruments whose measurements are often used as ground truth to validate optical properties inferred from ocean-color satellite observations. They therefore require high accuracy, in the 1 % range. From Morrow et al. (2010) [40], the average unbiased percentage difference between the C-OPS calibrated instrument and the SuBOPS (Submersible Biospherical Optical Profiling System) reference for a subset of 9 spectral channels was 1.8 %. This ±2.5 % uncertainty also includes possible drifts of the C-OPS in the range of 1-2% per year (instrument calibrated 19 months prior our usage).

**Table 1. Effective spectral-radiance calibration coefficients $R_i^{-1}$ for each spectral band $i$ calculated using Eq. (9). The exposure time and ISO x 0.01 gain of the calibration images are respectively 0.1 s and 1.0.**

| Spectral band $i$ | $\overline{L}_{i,source}$ [mW·sr$^{-1}$·m$^{-2}$·nm$^{-1}$] | $y_{DN,i,source}$ [x 10$^3$ ADU·s$^{-1}$] | $R_i^{-1}$ [W·sr$^{-1}$·m$^{-2}$·nm$^{-1}$·ADU$^{-1}$·s] | Uncertainty [%] |
|---|---|---|---|---|
| $r$ | $0.74 \pm 0.08$ | $61 \pm 1$ | 1.2 x 10$^{-8}$ | 10.4 |
| $g$ | $0.48 \pm 0.05$ | $65 \pm 1$ | 7.5 x 10$^{-9}$ | 10.6 |
| $b$ | $0.26 \pm 0.03$ | $21.9 \pm 0.5$ | 1.2 x 10$^{-8}$ | 11.0 |

### *3.6 Immersion factor*

Figure 8 shows the depth-dependent natural logarithm of digital numbers below water and the extrapolated $\ln y_{DN,i}(0^-)$. The in-air measurements are given in Table 2. As pointed out by Zibordi (2006), the need to correct for absorption ($a$ [m$^{-1}$]) and scattering ($b$ [m$^{-1}$]) of pure water – light attenuation following $e^{-(a+b)z}$ – is largely relaxed by using the logarithm of the $y_{DN,i}(z)$. A relatively good linear fit of the log-transformed digital numbers as a function of depth was obtained with $R^2$ values over 0.75 (see Fig. 8). Fluctuations in the measurements at a certain depth can result from small surface wavelets or an accumulation of bubbles on the optic external interface. Table 2 gives the immersion factors and relevant quantities to compute them using Eq. (11). We found immersion factors of 1.69 ± 0.02, 1.69 ± 0.02 and 1.71 ± 0.03 for the red, green, and blue bands. The increase in radiance below water (n-squared law $t_{aw} \cdot n_w$) was calculated using the refractive index of pure water from Hale and Querry (1973) [62] at the effective wavelength $\lambda_{eff}$ of the wavebands. We also used Fresnel equations for the transmittance. The fitted immersion factors of the three channels are equivalent inside their uncertainties due to little dispersion in visible light of the involved media and the large overlapping spectral bands.

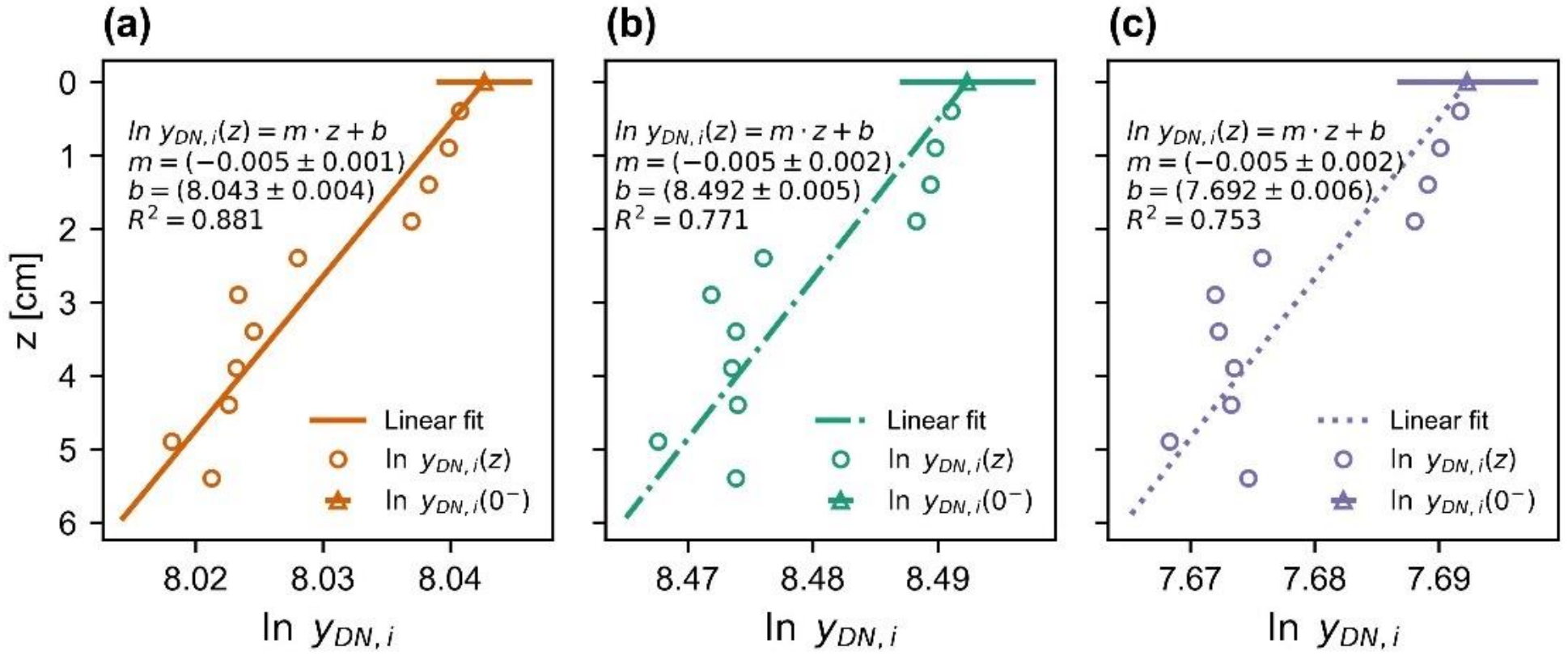

Fig. 8. Immersion factor characterization in the (a) red, (b) green and (c) blue wavebands. The circle markers represent the natural logarithm of the averaged in-water $y_{DN,i}$ measurements as a function of water height $z$ above the camera and corrected for n-squared law. The linear regression results are shown with the intercept $b$ corresponding to the camera response just below water (triangle markers).

Using the refractive index of pure water ($n_w$) at 25 °C at the effective wavelength of the spectral bands, we obtained external surface refractive indexes ($n_g$) of 1.60, 1.63, and 1.58 from the theoretical equation of immersion factor (see Eq. (10)) for the red, green, blue channels respectively. These values fall within the range of refractive indexes for various plastics. The uncertainties relating to $I_i$ as presented in Table 2, were calculated by a propagation of in-air digital-numbers uncertainty, which is the standard deviation of the values inside the averaging region and uncertainty relating to $y_{DN,i}(0^-)$ calculated by taking the 95 % confidence interval of the regression intercept standard error. From Eq. (10), the effect of salinity and temperature on the refractive index of water and therefore on $I_i$ was verified. Using the empirical equation of Quan and Fry (1995) [63] (valid for brine water [64]) for temperature $T$ between -2°C and 25°C and salinity $S$ between 0 and 35 PSU, it was found that the maximum variation in $I_i$ is an increase of approximately 0.02 from the value for pure water ($T$=25° C, $S$=0 PSU), which is in the same range as the uncertainties. Thus, no further correction was applied for variations in temperature and/or salinity.

**Table 2. Immersion factor results for each waveband. In order from left to right, the table shows the spectral bands, the measurements taken in-air $0^+$, the increase factor of radiance (n-squared law), the measurements just below water, $0^-$ , and the retrieved immersion factor. The measurements below water are corrected for the n-squared law as mentioned in the heading of the 4th column.**

| Spectral band $i$ | $y_{DN,i}(0^+)$ [ADU] | $t_{aw}n_w^2$ | $y_{DN,i}(0^-)/t_{aw}n_w^2$ [ADU] | Immersion factor |
|---|---|---|---|---|
| *r* | 5267 ± 68 | 1.738 | 3111 ± 12 | 1.69 ± 0.02 |
| *g* | 8259 ± 90 | 1.742 | 4877 ± 26 | 1.69 ± 0.02 |
| *b* | 3737 ± 60 | 1.748 | 2191 ± 12 | 1.71 ± 0.03 |

### *3.7 Dark frame*

Dark-frame characterization and analysis were performed to estimate electronic noise levels and determine whether some of these could be problematic in specific acquisition conditions (e.g. low light).

Figure 9(a) presents the histograms of the dark signal (15 frames pixel-wise average) at a minimum ISO of 100 and for each exposure time: 1/4000 s, 1/1000 s, 1/240 s, 1/60 s, 1/15 s, 1/5 s, 1 s, and 2 s. The gaussian-shaped curves do not vary significantly with exposure time at the lowest gain. Between 1/4000 s and 2 s, the central values of the histograms increase from 800.20 to 801.90 (over 16 384 possible counts) while the widths increase from 1.92 to 2.02. The central and width values are given in Fig. 9(c) and correspond respectively to the average and standard deviation of gaussian fits. This suggests that the temporal thermal dark current is rather low and that for the minimal gain setting, pixel bias (FPN) and reading noise mostly contribute to the dark frame. The dark current seems to start to become significant when grabbing times are in the order of seconds and over (see Fig. 9(c)). However, more exposure times would be required to clearly identify dark current as the cause. We noticed that the frames' average and standard deviations deviate from the fitted values when exposure time reaches 1/15 s and above. Histograms in Fig. 9(a) show only digital numbers between 790 and 810 as the majority of the 12 megapixels are in that range. Thus, the fit does not consider hot pixels that appear at high exposure-time settings. Careful attention should be taken to remove those hot pixels in low-light conditions, when $t_{\text{int}}$ goes over 1/15 s.

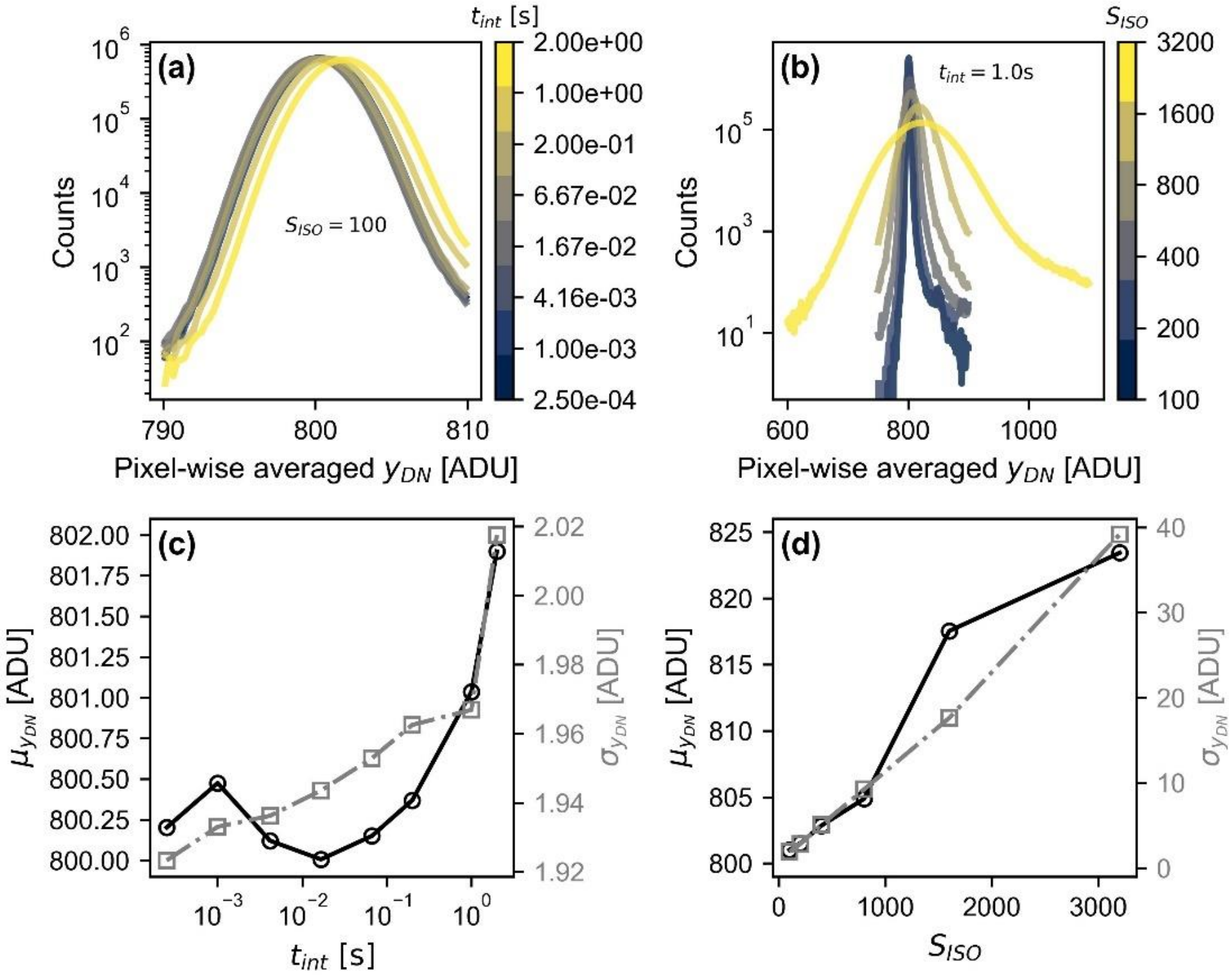

Fig. 9. Dark-frame characterization of the entire CMOS area (3456 x 3456 pixels) for multiple exposure conditions. The dark frames were analyzed using 15 captures at each condition. (a) Pixel dark-value histograms for ISO gain $S_{ISO}$ of 100 and multiple exposure times: 1/4000 s, 1/1000 s, 1/240 s, 1/60 s, 1/15 s, 1/5 s, 1 s and 2 s. (b) Dark counts for ISO-gain variation from 100 to 3200 by doubling the values each time and with exposure time of 1.0 s. The gaussian-fit central values and standard deviations of the dark images for exposure time and ISO-gain variations are presented in (c) and (d) respectively. For ISO 1600 and 3200 in figure (d), high noise levels across images caused the central values of the histograms (black points and curve) to deviate from the general trend, making these points appear as outliers.

Figure 9(b) shows the histograms of pixel counts for the pixel-wise averaged dark image of 15 frames for gain variations. We display a limiting case scenario with the exposure time at 1.0 s for each setting of ISO gain varied from 100 to 3200 by doubling each time. As expected, the dark image has more noise when the gain is increased. Both the average and standard deviations of the gaussian fit over the histograms increase close to linearly with ISO. From 100 to 3200, the average digital value increases from 801.02 to 823.43 and the standard deviation rises from 1.93 to 39.17. This is expected because the gain, which is assumed to be analog, amplifies both the signal and noise. For this reason, at low light, it is suggested that the exposure time be set initially, and then, if it is not enough, that the gain be increased. Furthermore, each time an acquisition with known exposure parameters is taken, it is recommended to take a dark frame – by covering the camera – at the same exposure time and gain to correct for the bias. Such dark frames should be taken in the same thermal conditions as frames exposed to light. If used in ISO-priority mode with unknown exposure times, an estimate of the bias level can be obtained by averaging the values in one of the sensor corners (see Fig. S5 in the supplemental document) that remain unexposed to light due to the circular fisheye image. This technique does not correct for the dark noise non-uniformity (DSNU) but can compensate any drifts in the bias.

### *3.8 Real-world validation*

Figure 10 shows the results of the real-world validation. The subfigures in the left column display the temporal measurements from the C-OPS radiometer (all data gathered at 6 Hz) and the Insta360® ONE radiance camera (taken every 2 minutes) for the red, green, and blue spectral channels. The C-OPS channels selected were those closest to the effective wavelength of each radiance camera band, and correspond respectively to the channels at 625, 555 and 490 nm. The radiance time series from the omnidirectional camera follow those of the scientific radiometer globally well. In the dynamic parts of the curves, discrepancies are observed. These likely result from the passage of clouds over both sensors (see Fig. S5 in the supplemental document). The error bars displayed for the radiance values of the 360-degree camera (see Fig. 10) also reflected their presence as they caused larger variations of the signal inside the 9˚ half FOV averaging region. Clouds can completely modify the spectral composition of the incoming light which, coupled with the camera's broad spectral bands (as discussed in Section 3.5), may cause greater errors in retrieved radiance. To a lesser extent, the gaps might be a consequence of timestamp mismatches between both instruments caused by unsynchronized clocks or differences in their temporal resolution.

In smoother regions – between 16h30 and 16h40 for example – there are better agreements between the two curves. This is more evident in the right column (Fig. 10(b), (d), and (f)), which compares measurements of the two sensors for a total of 30 data points. Some of these points, particularly in the lower part, tend to fall below the 1:1 line, especially for the 480-nm camera channel. Nonetheless, the fitted slopes – 1.0 ± 0.3 for the red and green bands, and 1.1 ± 0.3 for the blue channel – all include the 1:1 line within their uncertainty, indicating good data correspondence despite the large uncertainties. The larger slope in the blue channel might be explained by the broad camera bandwidths as the sky spectral radiance inside the spectral range of the band is less smooth than for the two other channels (verified by plotting the C-OPS radiance in its 19 spectral channels, see Fig. S4 in the supplemental document). We however observed the highest linear correlation with the C-OPS measurements for the same blue band. Its Pearson correlation coefficient is $r = 0.856$ ($R^2 = 0.732$), while $r = 0.843$ ($R^2 = 0.710$) at 540 nm and $r = 0.837$ ($R^2 = 0.700$) at 600 nm. The dispersion of the values from the linear fit is large with root-mean-square error (RMSE) in the range of 0.010, 0.011 and 0.012 $W{\cdot}sr^{-1}{\cdot}m^{-2}{\cdot}nm^{-1}$ for the red, green, and blue channels respectively. These values are high considering that most of the spectral radiance measured is within 0.12 $W{\cdot}sr^{-1}{\cdot}m^{-2}{\cdot}nm^{-1}$. The dispersion seems to increase as the values rise, but again this may result from the cloud presence and clock shifts during those measurements. We also computed the root-mean-square error directly between the sensor measurements. The results were slightly higher but rounded off to the same values as the RMSE from the linear fit: 0.010, 0.011 and 0.012 $W{\cdot}sr^{-1}{\cdot}m^{-2}{\cdot}nm^{-1}$ for the red, green, and blue channels respectively. This suggests that the comparison between both radiometers closely aligns with the 1:1 line, despite having large uncertainties. The median uncertainties of the 360-degree camera effective radiance values are 0.004 $W{\cdot}sr^{-1}{\cdot}m^{-2}{\cdot}nm^{-1}$ for the red and green bands, and 0.005 $W{\cdot}sr^{-1}{\cdot}m^{-2}{\cdot}nm^{-1}$ for the blue channel. For the mean unbiased percentage difference (MUPD), the values are 21.1 % for 600 nm, 20.3 % for 540 nm and 18.6 % for 480 nm. We also calculated the MUPD between the camera measurements and C-OPS values convolved with the camera spectral bands. We observed slight increase of the MUPD in each band to 29.2 %, 26.1 % and 22.2 % in the red, green, and blue channels respectively. The average percentage difference is lower for the 480 nm band, which may be because radiance in this spectral region is greater. If the absolute differences are not divided by the mean between the C-OPS and Insta360® measurements as done in Eq. (12), these differences are higher at 480 nm than in the other bands.

Other sources of error may explain the discrepancies between the two instruments. First, there are calibration errors as discussed in previous sections. These include uncertainties relating to the absolute spectral-radiance calibration coefficients (inverse radiance responsivities) and the fitted relative illumination. In addition, there are inter-pixel variations in sensitivity that were not accounted for. One way to reduce these sensitivity discrepancies would be to use a larger integrating sphere for uniform illumination and to apply a correction factor for each pixel illuminated. Another source of error is stray light which may impact the two radiometers at different levels, especially in clear sky conditions. There is also possible drift in the response of the C-OPS radiance sensor, as well as drift of the 360-degree camera since calibration. As discussed several times already, the broad bandwidths of the camera spectral responses are significant sources of bias. For future usage of the camera, we recommend conducting a simulation exercise using different spectra typical of the specific natural environment probed, in order to estimate the degree of errors introduced by the effective spectral radiance inside the broad bands. Overall, the agreements between the trusted commercial C-OPS and the calibrated omnidirectional radiance camera are in the range of what was expected for a consumer-grade camera. Similar inter-comparisons between RadCam fisheye [28] and Sea-Bird Scientific radiometers (HyperOCR and OCR-504I/R) were conducted with values of mean unbiased percentage differences of 16.3 % in laboratory and 6.2 % for upwelling radiance profiles inside the ocean. For commercial radiance sensors calibrated under the same conditions, the percentage of differences reach as low as ≤ 5 % over all the spectral points when compared with one other in the field [65,66].

It is important to note that the real-world validation experiment was performed in air, and the discrepancies observed cannot be fully generalized to in-water measurements. Indeed, the calibration pipeline for underwater radiance includes an additional immersion factor. Moreover, as presented in Section 3.3, the roll-off assessment in water has larger uncertainties compared to the in-air counterpart due to the inherently more complex method required to properly characterize the quantity in a wet medium. In future validations, real-world experiments in wet environments should be performed. The roll-off and immersion factor calibrations should be optimized to reduce their respective uncertainties.

The dynamic range of this real-world validation experiment is on one order of magnitude with spectral-radiance values between 0.0096-0.1207 $W \cdot sr^{-1} \cdot m^{-2} \cdot nm^{-1}$. We are however confident that with this 360-degree camera, we can obtain good radiance signal over 6 orders of magnitude ($10^{-6}$-$10^{0}$ $W \cdot sr^{-1} \cdot m^{-2} \cdot nm^{-1}$). From the linearity dataset of Section 3.1, it was possible to evaluate the signal-to-noise ratio (SNR) as a function of the dark-corrected digital numbers. By fitting the SNR curves according to the photon-transfer equations [67], we obtained noise-equivalent radiance (NER) signals – i.e. when the SNR is 1.0 – of 1.1 x $10^{-7}$, 4.5 x $10^{-8}$, and 1.1 x $10^{-7}$ $W \cdot sr^{-1} \cdot m^{-2} \cdot nm^{-1}$ for the 600, 540 and 480 nm bands respectively. For these calculations, we assumed an ISO gain of 100 and an exposure time of 1.0 s. We could get smaller NER by increasing the exposure parameters, but we kept conservative values of gain and exposure time to ensure limited impacts of hot pixels and thermal noises on the image. As a basis of comparison, Sea-Bird Scientific OCR-507, TriOS RAMSES ARC-VIS and Biospherical Instruments Inc. C-OPS radiometers have NER of 3.0 x $10^{-3}$, 2.5 x $10^{-7}$ (exposure time of 8.0 s, 500 nm), and 1.8 x $10^{-8}$ $W \cdot sr^{-1} \cdot m^{-2} \cdot nm^{-1}$ (at an acquisition rate of 5 Hz, 490 nm) respectively. A more reasonable limit of detection would be for a SNR of 3.0 which gives minimum measurable spectral radiances for the Insta360® ONE camera of 4.6 x $10^{-7}$, 1.8 x $10^{-7}$, and 4.6 x $10^{-7}$ $W \cdot sr^{-1} \cdot m^{-2} \cdot nm^{-1}$ for the red, green, and blue channels.

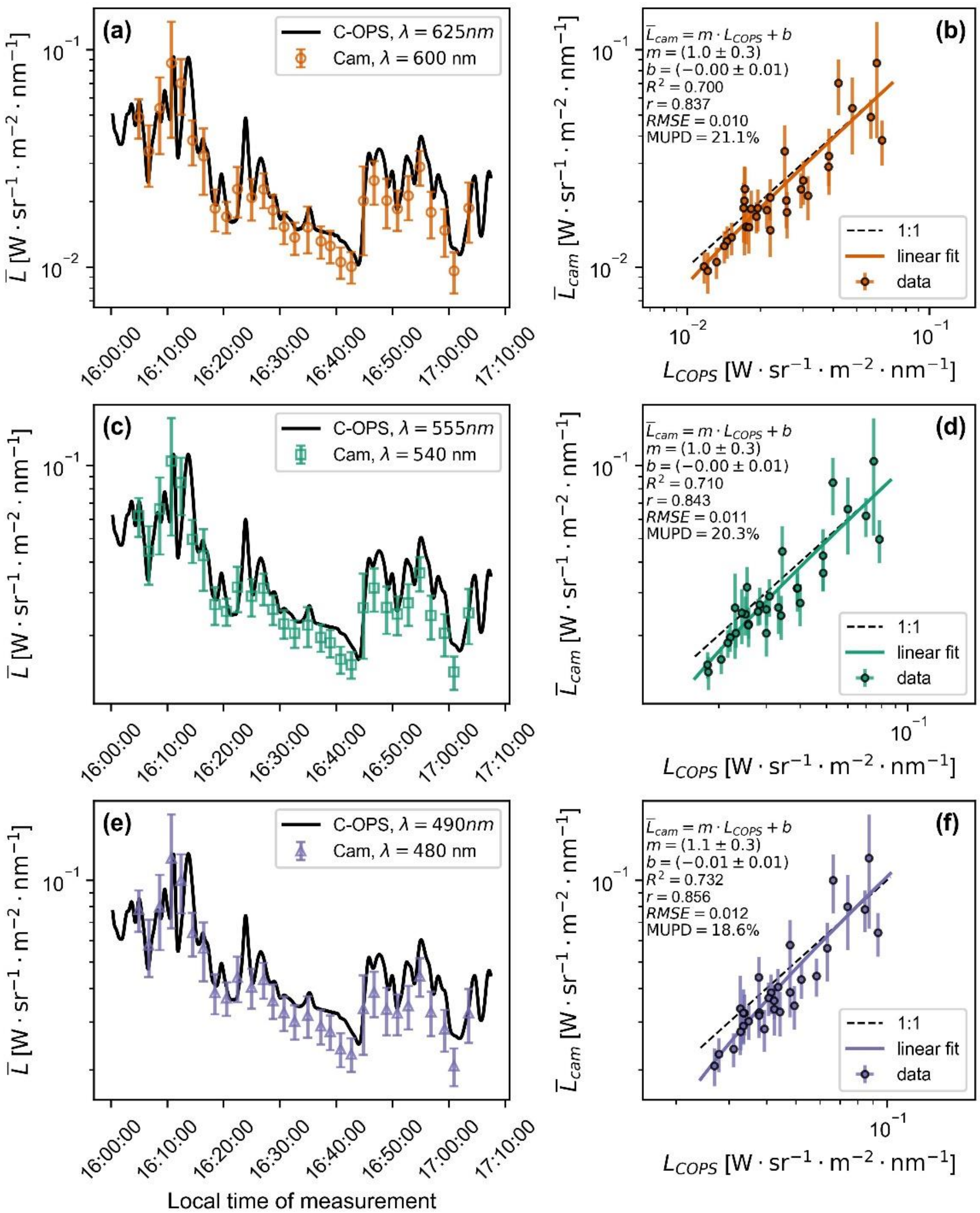

Fig. 10. Real-world validation experiment between the radiance camera and the commercial Compact Optical Profiling System (C-OPS) conducted on the Université Laval Campus in Quebec City. In the left column of the figure, the time series of the spectral radiance (units of W $sr^{-1}$ $m^{-2}$ $nm^{-1}$) from the sky light is shown for both instruments at three spectral bands: respectively red, green, and blue for (a), (b) and (c) subfigures. The right column shows scatterplots of camera versus C-OPS radiance measurements. The dotted line in each graphic is the 1:1 line, while the solid lines show the linear regressions. The uncertainties of the slope $m$ and intercept $b$ (given in each subfigure) reflect the 95 % confidence interval.

## 4. Conclusion

We successfully calibrated the Insta360® ONE 360-degree camera for in-air and in-water radiometry. The CMOS sensor response was assessed linear with exposure time with a mean Pearson correlation coefficient of 0.9992, and slightly less over gain with a value of 0.9961. For this reason, we suggest fixing a low gain for minimum noise amplification, and then allowing the camera to adjust its sensitivity with the exposure time. The projection functions of the optics follow linear relationships with field angles over FOV of 182˚, 152˚ in air and water respectively. The minimum detectable radiance levels for a SNR of 3.0 are $4.6 \times 10^{-7}$, $1.8 \times 10^{-7}$, and $4.6 \times 10^{-7}$ $W \cdot sr^{-1} \cdot m^{-2} \cdot nm^{-1}$ for the 600 nm, 540 nm, and 480 nm channels respectively. Satisfactory agreements were reached by comparing sky downward radiance measurements with simultaneous measurements of the C-OPS radiometer, which gave spectral MUPD between 18.6-21.1 %. We suspect that the broad spectral responses of the channels are the principal contributors to the observed discrepancies.

From these results, we confirm the potential of using raw images from the 360-degree camera for radiometry. To the best of our knowledge, this is the first time that an off-the-shelf 360-degree camera has been calibrated for radiometric/geometric absolute-light-field quantification. Since this type of camera is cost-efficient, compact, and easy-to-use, it can be easily adapted to a wide range of research needs. We believe that with proper calibration parameters stored in a database, these cameras could democratize the measurement of spectral radiance angular distributions and benefit numerous scientific and engineering projects. With their reduced form factors and low weight, light-field geometric information could be gathered from places such as unmanned aerial vehicles (UAV), AUV, or drilled holes inside solids (e.g. snow, sea ice). This would allow a broad diversity of measurements such as directional reflectance properties of multiple surfaces (sea ice, snow, ocean, clouds, etc.) over large spatial transects, vertical profiles of most apparent optical properties inside sea ice, or transmittance measurements below sea ice. Quantities thus captured would help increase understanding of the transfer of radiation inside those natural environments. The results of this study further support applications such as measurements of nondimensional geometric metrics like average cosines in different media, leaf area index in canopies, and cloud cover fraction in the sky. Moreover, the ratio between spectral band measurements could potentially be used as proxies for detecting the presence or concentration of chlorophyll.

Future real-world validations will involve field inter-comparisons over a wider dynamic range, under covered sky conditions – making the incident light field more isotropic –, and inside water. We also plan to perform temperature-dependency analysis. One way to reduce the impact of thermal noise is to acquire dark frames at the same temperature as the light frames. Future investigations will, moreover, encompass characterizations of both stray light and hot pixels. Quantification and spatial consistency across exposures of hot pixels will allow us to better understand and mitigate potential impacts on the applications discussed above. For further and more precise metrology usage, we emphasize on the need to calibrate for absolute radiance using a standard FEL lamp. The next steps also include testing the repeatability of the calibration for multiple ONE models to allow transferability of the calibration achieved in the present study. Moreover, new models of 360-degree devices with better specifications in terms of framerate, resolution, battery lifetime or sensitivity could also be added to the list of radiometrically calibrated omnidirectional cameras.

**Funding.** Raphaël Larouche was supported by the SMAART program through the Collaborative Research and Training Experience program (CREATE) of the Natural Sciences and Engineering Council of Canada (NSERC). This research was supported by the Sentinel North program of Université Laval, made possible, in part, thanks to the funding from the Canada First Research Excellence Fund, the Canadian Excellence Research Chair on Remote sensing of Canada's new Arctic frontier, and Marcel Babin Discovery Grant #RGPIN-2020-06384.

**Acknowledgments.** The authors would like to thank Anne-Sophie Poulin-Girard, Denis Brousseau, and Hugues Auger for assistance regarding material gathering and laboratory set-up for the calibrations/characterization methodologies. We also want to acknowledge Gabriel Lachance from Sophie LaRochelle's group that generously shared the integrating sphere. We also thank Marie-Hélène Forget for all her support as well as all members of the LRIO research group and Takuvik optical-engineering and instrumentation team (Yasmine Alikacem, Christophe Perron, Beatrice Lessard-Hamel, Bastian Raulier, Guislain Bécu, Eric Rhem) for fruitful discussions pertaining to this work. We gratefully acknowledge the scientific and financial support of ArcticNet and Québec-Océan.

**Disclosures.** The authors declare no conflicts of interest.

**Data availability.** Scripts for the calibration process are available on Github, https://github.com/RaphaelLarouche/radiance_camera_insta360, as well as in the following online repository DOI: https://doi.org/10.5281/zenodo.4660993. The calibration dataset consisting of raw images can be found in the Zenodo repository DOI: https://doi.org/10.5281/zenodo.10278731.

**Supplemental document.** See Supplement 1 for supporting content.

**References**

1. C. Mobley, *Light and Water: Radiative Transfer in Natural Waters* (1994).
2. D. Vansteenwegen, K. Ruddick, A. Cattrijsse, Q. Vanhellemont, and M. Beck, "The Pan-and-Tilt Hyperspectral Radiometer System (PANTHYR) for Autonomous Satellite Validation Measurements—Prototype Design and Testing," Remote Sens (Basel) **11**, 1360 (2019).
3. K. Ruddick, V. De Cauwer, Y. Park, G. Becu, J.-P. Blauwe, E. Vreker, P.-Y. Deschamps, M. Knockaert, B. Nechad, A. Pollentier, P. Roose, D. Saudemont, and D. Tuyckom, "Preliminary validation of MERIS water products for Belgian coastal waters," (2002).
4. P. Nandy, K. Thome, and S. Biggar, "Instrument for retrieval of BRDF data for vicarious calibration," in *IGARSS '98. Sensing and Managing the Environment. 1998 IEEE International Geoscience and Remote Sensing. Symposium Proceedings. (Cat. No.98CH36174)* (IEEE, 1998), pp. 562–564 vol.2.
5. K. J. Voss and A. Morel, "Bidirectional reflectance function for oceanic waters with varying chlorophyll concentrations: Measurements versus predictions," Limnol Oceanogr **50**, 698–705 (2005).
6. C. Goyens, S. Marty, E. Leymarie, D. Antoine, M. Babin, and S. Bélanger, "High Angular Resolution Measurements of the Anisotropy of Reflectance of Sea Ice and Snow," Earth and Space Science **5**, 30–47 (2018).
7. T. Carlsen, G. Birnbaum, A. Ehrlich, V. Helm, E. Jäkel, M. Schäfer, and M. Wendisch, "Parameterizing anisotropic reflectance of snow surfaces from airborne digital camera observations in Antarctica," Cryosphere **14**, 3959–3978 (2020).
8. A. Jechow, Z. Kolláth, S. J. Ribas, H. Spoelstra, F. Hölker, and C. C. M. Kyba, "Imaging and mapping the impact of clouds on skyglow with all-sky photometry," Sci Rep **7**, 6741 (2017).
9. A. Jechow, C. Kyba, and F. Hölker, "Beyond All-Sky: Assessing Ecological Light Pollution Using Multi-Spectral Full-Sphere Fisheye Lens Imaging," J Imaging **5**, 46 (2019).
10. Z. Kolláth, A. Cool, A. Jechow, K. Kolláth, D. Száz, and K. P. Tong, "Introducing the dark sky unit for multi-spectral measurement of the night sky quality with commercial digital cameras," J Quant Spectrosc Radiat Transf **253**, 107162 (2020).
11. K. Parpairi, N. Baker, K. Steemers, and R. Compagnon, "The Luminance Differences index: a new indicator of user preferences in daylit spaces," Lighting Research & Technology **34**, 53–66 (2002).
12. P. Lalande, C. M. H. Demers, J.-F. Lalonde, A. Potvin, and M. Hébert, "Spatial representations of melanopic light in architecture," Archit Sci Rev **64**, 522–533 (2021).
13. K. J. Voss, "Use of the radiance distribution to measure the optical absorption coefficient in the ocean," Limnol Oceanogr **34**, 1614–1622 (1989).
14. M. R. Lewis, J. Wei, R. Van Dommelen, and K. J. Voss, "Quantitative estimation of the underwater radiance distribution," J Geophys Res Oceans **116**, (2011).
15. D. Antoine, A. Morel, E. Leymarie, A. Houyou, B. Gentili, S. Victori, J.-P. Buis, N. Buis, S. Meunier, M. Canini, D. Crozel, B. Fougnie, and P. Henry, "Underwater Radiance Distributions Measured with Miniaturized Multispectral Radiance Cameras," J Atmos Ocean Technol **30**, 74–95 (2013).
16. A. Morel, K. J. Voss, and B. Gentili, "Bidirectional reflectance of oceanic waters: A comparison of modeled and measured upward radiance fields," J Geophys Res **100**, 13143–13150 (1995).

17. D. K. Perovich, C. S. Roesler, and W. S. Pegau, "Variability in Arctic sea ice optical properties," J Geophys Res Oceans **103**, 1193–1208 (1998).
18. W. S. Pegau and J. R. V Zaneveld, "Field measurements of in-ice radiance," Cold Reg Sci Technol **31**, 33–46 (2000).
19. Y. Qin, M. A. Box, and D. L. B. Jupp, "Inversion of multiangle sky radiance measurements for the retrieval of atmospheric optical properties 1. Algorithm," J Geophys Res **107**, 4652 (2002).
20. W. Caldwell and V. C. Vanderbilt, "Tree Canopy Radiance Measurement System," Optical Engineering **28**, 281227 (1989).
21. A. Gershun, "The Light Field," Journal of Mathematics and Physics **18**, 51–151 (1939).
22. J. E. (John E. Tyler, *Design of an Underwater Radiance Photometer* , Report ; No. 3-2 (S.Q. Duntley, 1959).
23. R. C. Smith, R. W. Austin, and J. E. Tyler, "An Oceanographic Radiance Distribution Camera System," Appl Opt **9**, 2015 (1970).
24. K. J. Voss and G. Zibordi, "Radiometric and Geometric Calibration of a Visible Spectral Electro-Optic “Fisheye” Camera Radiance Distribution System," J Atmos Ocean Technol **6**, 652–662 (1989).
25. K. J. Voss and A. L. Chapin, "Next-generation in-water radiance distribution camera system," in *Ocean Optics XI*, G. D. Gilbert, ed. (1992), Vol. 1750, pp. 384–387.
26. K. J. Voss and A. L. Chapin, "Upwelling radiance distribution camera system, NURADS," Opt Express **13**, 4250 (2005).
27. K. J. Voss, A. Morel, and D. Antoine, "Detailed validation of the bidirectional effect in various Case 1 waters for application to ocean color imagery," Biogeosciences **4**, 781–789 (2007).
28. J. Wei, R. Van Dommelen, M. R. Lewis, S. McLean, and K. J. Voss, "A new instrument for measuring the high dynamic range radiance distribution in near-surface sea water," Opt Express **20**, 27024–27038 (2012).
29. M. Darecki, D. Stramski, and M. Sokólski, "Measurements of high-frequency light fluctuations induced by sea surface waves with an Underwater Porcupine Radiometer System," J Geophys Res Oceans **116**, (2011).
30. S. Thibault, J. Parent, H. Zhang, and P. Roulet, "Design, Fabrication and Test of Miniature Plastic Panomorph Lenses with 180° Field of View," in *Classical Optics 2014* (Optica Publishing Group, 2014), p. IM2A.3.
31. B. Light, T. C. Grenfell, and D. K. Perovich, "Transmission and absorption of solar radiation by Arctic sea ice during the melt season," J Geophys Res **113**, C03023 (2008).
32. J. K. Ehn, T. N. Papakyriakou, and D. G. Barber, "Inference of optical properties from radiation profiles within melting landfast sea ice," J Geophys Res **113**, C09024 (2008).
33. Z. Xu, Y. Yang, Z. Sun, Z. Li, W. Cao, and H. Ye, "In situ measurement of the solar radiance distribution within sea ice in Liaodong Bay, China," Cold Reg Sci Technol **71**, 23–33 (2012).
34. B. Light, D. K. Perovich, M. A. Webster, C. Polashenski, and R. Dadic, "Optical properties of melting first-year Arctic sea ice," J Geophys Res Oceans **120**, 7657–7675 (2015).
35. A. Ehrlich, E. Bierwirth, M. Wendisch, A. Herber, and J.-F. Gayet, "Airborne hyperspectral observations of surface and cloud directional reflectivity using a commercial digital camera," Atmos Chem Phys **12**, 3493–3510 (2012).
36. S. Becker, A. Ehrlich, E. Jäkel, T. Carlsen, M. Schäfer, and M. Wendisch, "Airborne measurements of directional reflectivity over the Arctic marginal sea ice zone," Atmos Meas Tech **15**, 2939–2953 (2022).
37. D. S. Kimes, "Dynamics of directional reflectance factor distributions for vegetation canopies," Appl Opt **22**, 1364–1372 (1983).
38. L. Li, X. Mu, J. Qi, J. Pisek, P. Roosjen, G. Yan, H. Huang, S. Liu, and F. Baret, "Characterizing reflectance anisotropy of background soil in open-canopy plantations using

UAV-based multiangular images," ISPRS Journal of Photogrammetry and Remote Sensing **177**, 263–278 (2021).

39. A. Lorente, K. F. Boersma, P. Stammes, L. G. Tilstra, A. Richter, H. Yu, S. Kharbouche, and J.-P. Muller, "The importance of surface reflectance anisotropy for cloud and NO2 retrievals from GOME-2 and OMI," Atmos Meas Tech **11**, 4509–4529 (2018).
40. J. H. Morrow, S. B. Hooker, C. R. Booth, G. Bernhard, R. N. Lind, and J. W. Brown, "Advances in measuring the apparent optical properties (AOPs) of optically complex waters," NASA Tech. Memo **215856**, 42–50 (2010).
41. R. D. Fiete, *Modeling the Imaging Chain of Digital Cameras* (SPIE, 2010).
42. N. Hagen, "Flatfield correction errors due to spectral mismatching," Optical Engineering **53**, 123107 (2014).
43. M. Pagnutti, R. E. Ryan, G. Cazenavette, M. Gold, R. Harlan, E. Leggett, and J. Pagnutti, "Laying the foundation to use Raspberry Pi 3 V2 camera module imagery for scientific and engineering purposes," J Electron Imaging **26**, 013014 (2017).
44. T. P. Johnson and J. Sasian, "Image distortion, pupil coma, and relative illumination," Appl Opt **59**, G19–G23 (2020).
45. G. Zibordi, "Immersion Factor of In-Water Radiance Sensors: Assessment for a Class of Radiometers," J Atmos Ocean Technol **23**, 302–313 (2006).
46. J. R. Schott, *Remote Sensing: The Image Chain Approach* (Oxford University Press on Demand, 2007).
47. O. Burggraaff, N. Schmidt, J. Zamorano, K. Pauly, S. Pascual, C. Tapia, E. Spyrakos, and F. Snik, "Standardized spectral and radiometric calibration of consumer cameras," Opt Express **27**, 19075–19101 (2019).
48. Z. Wang, Z. Cen, and X. Li, "Two distortion correcting methods for fisheye images," in *AOPC 2017: Optical Sensing and Imaging Technology and Applications*, Y. Jiang, H. Gong, W. Chen, and J. Li, eds. (SPIE, 2017), p. 77.
49. D. Scaramuzza, A. Martinelli, and R. Siegwart, "A Toolbox for Easily Calibrating Omnidirectional Cameras," in *2006 IEEE/RSJ International Conference on Intelligent Robots and Systems* (IEEE, 2006), pp. 5695–5701.
50. D. Scaramuzza, A. Martinelli, and R. Siegwart, "A Flexible Technique for Accurate Omnidirectional Camera Calibration and Structure from Motion," in *Fourth IEEE International Conference on Computer Vision Systems (ICVS'06)* (IEEE, 2006), pp. 45–45.
51. G. Bradski, "The OpenCV Library," Dr. Dobb's Journal of Software Tools (2000).
52. Wonpil Yu, Yunkoo Chung, and Jung Soh, "Vignetting distortion correction method for high quality digital imaging," in *Proceedings of the 17th International Conference on Pattern Recognition, 2004. ICPR 2004.* (IEEE, 2004), pp. 666-669 Vol.3.
53. F. Sigernes, M. Dyrland, N. Peters, D. A. Lorentzen, T. Svenøe, K. Heia, S. Chernouss, C. S. Deehr, and M. Kosch, "The absolute sensitivity of digital colour cameras," Opt Express **17**, 20211–20220 (2009).
54. T. Suzuki, Y. Amano, J. Takiguchi, T. Hashizume, S. Suzuki, and A. Yamaba, "Development of low-cost and flexible vegetation monitoring system using small unmanned aerial vehicle," in *2009 ICCAS-SICE* (2009), pp. 4808–4812.
55. E. Berra, S. Gibson-Poole, A. MacArthur, R. Gaulton, and A. Hamilton, "Estimation of the spectral sensitivity functions of un-modified and modified commercial off-the-shelf digital cameras to enable their use as a multispectral imaging system for UAVs," The International Archives of the Photogrammetry, Remote Sensing and Spatial Information Sciences **XL-1/W4**, 207–214 (2015).
56. T. Leeuw and E. Boss, "The HydroColor App: Above Water Measurements of Remote Sensing Reflectance and Turbidity Using a Smartphone Camera," Sensors **18**, 256 (2018).

57. J.-H. Lee, W.-S. Choi, K.-H. Lee, and J.-B. Yoon, "A simple and effective fabrication method for various 3D microstructures: backside 3D diffuser lithography," Journal of Micromechanics and Microengineering **18**, 125015 (2008).
58. A. Sciandra, L. Lazzara, H. Claustre, and M. Babin, "Responses of growth rate, pigment composition and optical properties of Cryptomonas sp. to light and nitrogen stresses," Mar Ecol Prog Ser **201**, 107–120 (2000).
59. F. Wang and A. Theuwissen, "Linearity analysis of a CMOS image sensor," Electronic Imaging **29**, 84–90 (2017).
60. S. van der Walt, J. L. Schönberger, J. Nunez-Iglesias, F. Boulogne, J. D. Warner, N. Yager, E. Gouillart, and T. Yu, "scikit-image: image processing in Python," PeerJ **2**, e453 (2014).
61. Z. Zhuang, X. Dallaire, J. Parent, P. Roulet, and S. Thibault, "Wide angle lens with improved relative illumination characteristics," in *Current Developments in Lens Design and Optical Engineering XXI*, R. B. Johnson, V. N. Mahajan, and S. Thibault, eds. (SPIE, 2020), Vol. 11482, p. 114820E.
62. G. M. Hale and M. R. Querry, "Optical Constants of Water in the 200-nm to 200-μm Wavelength Region," Appl Opt **12**, 555–563 (1973).
63. X. Quan and E. S. Fry, "Empirical equation for the index of refraction of seawater," Appl Opt **34**, 3477 (1995).
64. J. R. Frisvad, "Empirical formula for the refractive index of freezing brine," Appl Opt **48**, 2149–2153 (2009).
65. V. Vabson, J. Kuusk, I. Ansko, R. Vendt, K. Alikas, K. Ruddick, A. Ansper, M. Bresciani, H. Burmester, M. Costa, D. D'Alimonte, G. Dall'Olmo, B. Damiri, T. Dinter, C. Giardino, K. Kangro, M. Ligi, B. Paavel, G. Tilstone, R. Van Dommelen, S. Wiegmann, A. Bracher, C. Donlon, and T. Casal, "Field Intercomparison of Radiometers Used for Satellite Validation in the 400–900 nm Range," Remote Sens (Basel) **11**, 1129 (2019).
66. G. Tilstone, G. Dall'Olmo, M. Hieronymi, K. Ruddick, M. Beck, M. Ligi, M. Costa, D. D'Alimonte, V. Vellucci, D. Vansteenwegen, A. Bracher, S. Wiegmann, J. Kuusk, V. Vabson, I. Ansko, R. Vendt, C. Donlon, and T. Casal, "Field Intercomparison of Radiometer Measurements for Ocean Colour Validation," Remote Sens (Basel) **12**, 1587 (2020).
67. J. R. Janesick, *Photon Transfer* (SPIE, 2007).